\documentclass[x11names,12pt,a4paper]{article}
\usepackage[margin=.85in]{geometry}
\usepackage{jheppub} 

\makeatletter
\renewcommand\section{\@startsection {section}{1}{\z@}%
                                 {-3.5ex \@plus -1ex \@minus -.2ex}
                                   {2.3ex \@plus.2ex}%
                                   {\normalfont\large\bfseries}}
\renewcommand\subsection{\@startsection{subsection}{2}{\z@}%
                                   {-3.25ex\@plus -1ex \@minus -.2ex}%
                                     {1.5ex \@plus .2ex}%
                                     {\normalfont\bfseries}}
\renewcommand\subsubsection{\@startsection{subsubsection}{3}{\z@}%
                                   {-3.25ex\@plus -1ex \@minus -.2ex}%
                                     {1.5ex \@plus .2ex}%
                                     {\normalfont\itshape}}
\makeatother

\usepackage{titlesec}

\usepackage[utf8]{inputenc}
\usepackage{lmodern}
\usepackage[T1]{fontenc} 
\usepackage{microtype} 

\usepackage{xcolor}
\definecolor{dark-red}{rgb}{0.7,0,0}
\definecolor{dark-green}{rgb}{0.1,0.4,0}
\definecolor{NiceBlue}{rgb}{0.30196,0.55294,0.57647}

\usepackage{hyperref}
\hypersetup{
      colorlinks=true,
      linkcolor=dark-green,
      citecolor=dark-red,
      urlcolor=dark-green,
}
\newcommand{\bea}{\begin{eqnarray}}
\newcommand{\eea}{\end{eqnarray}}

\usepackage{mathtools}
\usepackage{enumitem}

\def\bra{\langle}\def\ket{\rangle}

\newcommand{\dd}{\mathrm{d}}

\begin{document}
\begin{titlepage}
  \thispagestyle{empty}

\vspace*{-0.8cm}
\begin{center}

{\bf {\LARGE \bfseries  Spacetime from Operator Algebras}}

\begin{center}

\vspace{0.5cm}

{\bf Vyshnav Mohan \textbf{and} L\'arus Thorlacius}

\vspace{0.2cm}
  
\hspace{.03em}Science Institute,
University of Iceland \\Dunhaga 3, 107 Reykjav\'{i}k, Iceland.

\vspace{0.2cm}

{\tt \href{mailto:vyshnav.vijay.mohan@gmail.com}{vyshnav.vijay.mohan@gmail.com}, \href{mailto:lth@hi.is}{lth@hi.is} }
\end{center}

\vspace{1.2cm}
{\bf Abstract}
\end{center}
\begin{quotation}
\noindent Under suitable assumptions, geometric objects such as the spacetime metric and curvature tensor can be reconstructed from the algebra of operators of quantized matter fields in the limit of vanishing Newton's constant. In this framework, the full non-linear Einstein equations can be expressed in the language of operator algebras, extending Jacobson's derivation without invoking the area law for Bekenstein-Hawking entropy. These assumptions can then be used as a criterion for determining whether the semiclassical limit of a given quantum theory admits an emergent gravitational description. Going in the other direction, the discrete spectrum of a holographic theory at finite $N$ can be modelled by adding non-perturbative corrections to semiclassical operator algebras. The type III von Neumann algebra that arises in the vanishing Newton's constant limit can be enlarged by adjoining its modular Hamiltonian. A random matrix theory completion of this enlarged algebra, followed by ensemble averaging, results in a type I von Neumann algebra whose minimal projectors approximate those of the underlying microstates. In the case of an eternal black hole, the dimension of the type I algebra equals the Bekenstein-Hawking entropy with universal logarithmic corrections. The complexity of probe operators in the boundary theory provides a diagnostic of the validity of the corresponding bulk semiclassical effective field theory.

\end{quotation}
\vfill 

\end{titlepage}

\setcounter{page}{0}
\setcounter{tocdepth}{2}
\setcounter{footnote}{0}

\newpage
\pagestyle{plain}
\parskip 0.1in

\setcounter{page}{2}

\setcounter{tocdepth}{2}
\tableofcontents
\afterTocSpace
\hrule
\afterTocRuleSpace

\section{Introduction}
\label{introsec}
A profound realization of the past few decades is that gravity may be an emergent phenomenon arising from non-gravitational degrees of freedom \cite{Jacobson:1995ab,Padmanabhan:2009vy,Verlinde:2010hp,VanRaamsdonk:2010pw,Maldacena:2013xja} (see \cite{Bousso:2022ntt} for a more comprehensive list of references). From this perspective, the technical difficulties associated with the canonical quantization of gravity can be avoided. The AdS/CFT correspondence \cite{Maldacena:1997re,Witten:1998qj,Gubser:1998bc} provides a concrete realization of this idea, in which a theory of gravity emerges in the semiclassical limit of a non-gravitational quantum field theory. This naturally raises a fundamental question: what is the appropriate language for understanding the emergence of gravity?

An important clue comes from the reconstruction theorem of Connes \cite{Connes1994,Connes1995}. Let $M$ be a compact, oriented Riemannian spin manifold. Connes' reconstruction theorem states that such a manifold can be recovered from the commutative spectral triple
\bea
(C^{\infty}(M), \mathcal{H}, \mathcal{D})\,,
\eea
where $C^{\infty}(M)$ is the algebra of smooth complex-valued functions on $M$, $\mathcal{H} = L^2(M,S)$ is the Hilbert space of square-integrable sections of the spinor bundle $S \to M$, and $\mathcal{D}$ is the self-adjoint Dirac operator associated with the Levi-Civita connection. The algebra $C^{\infty}(M)$ acts on $\mathcal{H}$ by pointwise multiplication, and under suitable regularity conditions \cite{Connes1994}, the spectrum of the algebra recovers the underlying topological space, while the metric is reconstructed via Connes' distance formula. This reformulation of geometry follows a standard mathematical strategy where one replaces a complicated geometric object with the vector space of functions defined on it, and then uses linear algebra to study its properties (see \cite{Vogan2005} for further examples). In physics, this idea of reconstructing the geometry from its spectral properties is often referred to as the problem of ``hearing the shape of a drum'' \cite{Kac1966}.

When the goal is to understand the emergence of gravity from an underlying quantum theory, a natural extension of Connes' strategy is to study the algebra of operators acting on the Hilbert space of the theory itself. One would then like spacetime geometry and gravitational dynamics to emerge from this algebraic structure in an appropriate semiclassical limit. An indication that this is the correct framework comes from the $G_N \to 0$ limit of AdS/CFT, which corresponds to sending both $g_s$ and $\alpha'/L_{\text{AdS}}^2 $ to zero.  When a holographic dual exists, the corresponding boundary limit is given by $N,\lambda \to \infty$, where $\lambda$ is the 't Hooft coupling. In this regime, the bulk theory reduces to free fields propagating on a fixed background geometry.  Remarkably, geometric structures such as Killing horizons can already be seen to emerge purely from the operator algebra of the quantized matter fields \cite{Leutheusser:2021frk,Leutheusser:2021qhd,Ouseph:2023juq}, bolstering the intuition that operator algebras provide an appropriate framework for understanding the emergence of spacetime geometry.

In Section \ref{geometrysection}, we extend these results by showing that the operator algebra of the quantized matter fields encodes \emph{all} the local geometric information of the spacetime, not just that of the Killing horizons. In particular, we consider the $G_N \to 0$ limit, where the bulk description consists of free fields on some $D=d+1$-dimensional fixed curved manifold. For simplicity, we restrict the matter sector to a free massive scalar field. Such a description naturally arise in large $N$ holographic theories, for example in the bulk dual of $\mathcal{N}=4$ super Yang-Mills theory \cite{Leutheusser:2021frk,Leutheusser:2021qhd,Witten:2021unn}. Then, we show that the geometric data of the spacetime, such as the metric and curvature tensors, can be systematically extracted from the triple
\bea
(\mathcal{A}, \mathcal{H}, |\omega\rangle)\,,
\eea
where $\mathcal{A}$ is the operator algebra of the quantized scalar field, $\mathcal{H}$ is the Hilbert space on which these operators act, and $|\omega\rangle$ is some preferred `vacuum' vector in this Hilbert space, provided the following conditions are satisfied:
\begin{enumerate}[label=(\arabic*)]
\item \phantomsection\label{item:first}$|\omega\rangle$ satisfies the Hadamard condition.
\item \phantomsection\label{item:second}The algebra $\mathcal{A}$ contains a family of subalgebras $\mathcal{W}_j$ which satisfy certain conditions that we will specify in Section \ref{geometrysection}. We refer to these as \emph{local Rindler algebras}.
\item \phantomsection\label{item:third}The correlation functions of quantized scalar field in $|\omega\rangle$ has a well-defined large mass limit.
\end{enumerate}

Precise definitions are given in Section \ref{geometrysection}, but it is useful to briefly consider the physical content of these assumptions. Taken together, the two conditions \hyperref[item:first]{(1)}-\hyperref[item:second]{(2)} provide an operator algebraic realization of the equivalence principle, while condition \hyperref[item:third]{(3)} implements a semiclassical limit. 
The equivalence principle is not merely the statement that locally flat coordinates exist, since \emph{any} Lorentzian manifold admits such coordinate patches by definition. Rather, the crucial requirement is that one can pass to a local inertial frame in which non-gravitational physics reduces to special relativity up to tidal corrections. The Hadamard condition ensures that the short-distance behaviour of the scalar two-point function is universally governed by the flat space singularity structure. Local Rindler algebras are obtained by restricting the scalar field to Rindler wedges within local patches of the manifold. We then require that each such algebra can be approximated by an exact Rindler algebra. In Section \ref{geometrysection}, we show that this assumption can be used to identify the approximate local Poincar\'{e} generators of the spacetime. In this sense, the matter fields ``see'' a flat spacetime in the neighbourhood of every point.

Although we arrive at the conditions \hyperref[item:first]{(1)}-\hyperref[item:third]{(3)} by working with a scalar field living on a pre-existing geometry, the crucial point is that the conditions themselves are purely algebraic and do not require geometric input. This opens the door to the inverse problem. Given an abstract algebraic triple $(\mathcal{A}, \mathcal{H}, |\omega\rangle)$ arising as an appropriate semiclassical limit of some quantum theory, one can ask directly whether the conditions \hyperref[item:first]{(1)}-\hyperref[item:third]{(3)} are satisfied in the operator algebra, without assuming any background geometry. If they are, a spacetime geometry can be systematically reconstructed from the algebraic data alone. 

It remains unclear, however, whether the emergent geometry satisfies Einstein's field equations. In Section \ref{geometrysection}, we show that upon including perturbative $G_N$ corrections around the $G_N \to 0$ limit, additional assumptions guarantee the validity of the Einstein equations. In particular, we show that if, for every local Rindler wedge algebra $\mathcal{W}_j$, there exists a state $\omega_j$, which we refer to as a \emph{locally stationary state}, then the emergent geometry satisfies the full nonlinear Einstein equations sourced by the matter fields. 
There is a simple way to understand this condition. Jacobson showed how the Einstein equations can be derived from a local equilibrium condition imposed on the local Rindler horizons of the theory \cite{Jacobson:1995ab}. In the operator-algebraic approach, the existence of locally stationary states $\omega_j$ plays the role of this local equilibrium. There is, however, a crucial difference between Jacobson's argument and our result. Jacobson's derivation relies on the Bekenstein-Hawking entropy formula. In our approach this input is replaced by the existence of locally stationary states and their semiclassical coherent excitations, providing a novel perspective on the quantum origins of the Einstein equations. 

To summarize, the emergence of gravity can be reformulated in terms of an operator algebra, and associated Hilbert space, satisfying conditions \hyperref[item:first]{(1)}-\hyperref[item:third]{(3)}, together with the existence of the locally stationary states $\omega_j$. This characterization is independent of the spacetime dimension and applies to spacetimes with arbitrary asymptotics.

In most cases, however, the full quantum theory of gravity is not known, and a more useful question is whether one can instead work within the semiclassical theory and incorporate non-perturbative quantum gravity effects to model features of the full theory.\footnote{In this paper, we use the term \emph{semiclassical theory} to refer to the perturbative quantum gravity regime, in which we study the $G_N \to 0$ limit, together with perturbative $G_N$ corrections around it, but without non-perturbative effects in $G_N$.} In Section \ref{typeIsection}, we show that operator algebras provide a natural setting to incorporate these effects. 
To make concrete statements, we use holography. In particular, we study an eternal AdS black hole dual to a thermofield double state. The $G_N \to 0$ limit corresponds to the $N, \lambda \to \infty$ limit of the boundary theory. In this large $N$ limit, only single-trace operators survive, and the algebra of bulk scalar fields is dual to the algebra generated by these operators. It turns out that the algebra associated with a single boundary CFT is a type III von Neumann algebra \cite{Leutheusser:2021qhd,Leutheusser:2021frk,Furuya:2023fei}. This algebra is pathological in the sense that it admits neither density matrices nor a trace.

Including $1/N^2$ perturbative corrections around the $N \to \infty$ limit modifies the algebra to a type II$_\infty$ algebra \cite{Witten:2021unn}, allowing one to define a trace and density matrices. However, it still does not contain pure states. To access, for example, the microstates of the theory and count them, we need pure states. These states are only available in an algebra of type I. Unfortunately, there is no canonical way to obtain a type I algebra from a type III or type II algebra. This is not surprising, since there is no a priori reason for the semiclassical limit of a theory to encode information about the full quantum Hilbert space.

In Section \ref{typeiconstructionsection}, we show that there is nevertheless a natural way to obtain an \emph{approximate} type I algebra starting from the type II$_\infty$ algebra. A convenient way to systematically incorporate non-perturbative quantum gravity corrections is to treat the large $N$ perturbative quantum gravity limit as an effective random matrix theory (RMT) \cite{Banerjee:2024fmh,Magan:2024nkr}. Operationally, this amounts to replacing the density of states in the spectral decomposition of the elements of the type II$_\infty$ algebra by the RMT density of states.  We show that the resulting algebra can be ensemble averaged to produce a type I algebra acting on a finite-dimensional Hilbert space. In the case of an eternal black hole, the dimension of this Hilbert space is given by the exponential of the Bekenstein-Hawking entropy together with the universal logarithmic correction in the black hole area.

There are different ways to motivate this RMT construction, beyond the fact that it produces the desired algebraic structure. In the large $N$ limit, information about the discreteness of the energy spectrum is lost, and this discreteness is crucial for isolating individual microstates. The RMT construction introduces level repulsion, effectively emulating the spectral discreteness. 

The RMT description makes the chaotic nature of black holes manifest. This point makes it clear that the RMT construction is not intrinsically tied to gravity. The idea of using random matrix theory to describe universal spectral properties of complex quantum systems goes back to Wigner \cite{Wigner1951} and an emergent random matrix description in the semiclassical limit has been used in systems such as a gas of hard spheres in a box to understand thermalization and chaos \cite{Srednicki:1994mfb,Krishnan:2021faa}. From this perspective, the averaging and randomness in our construction provides a physically motivated approximation scheme for semiclassical theories, which is not specific to gravity. 

The finite-dimensional Hilbert space on which the type I algebra acts has recently been obtained in the literature from bulk wormhole contributions, using Euclidean path integral methods to construct approximate microstates \cite{Balasubramanian:2022lnw,Balasubramanian:2022gmo,Climent:2024trz}. Here, we sidestep the need for Euclidean path integrals, and instead proceed in close parallel with the algebraic computation of the generalized entropy of a black hole in \cite{Chandrasekaran:2022eqq}.

The paper is organized as follows. In Section \ref{geometrysection}, we show how the metric and curvature tensors can be extracted from operator algebras. In Section \ref{typeIsection}, we construct an approximate type I algebra starting from the type II algebra obtained by adjoining the modular Hamiltonian to the large $N$ type III algebra. 

\section{Emergence of Gravity in the Semiclassical Limit}
\label{geometrysection}
In this section, we work in the $G_N \to 0$ limit, where the theory reduces to free fields propagating on a fixed background. A key observation in \cite{Leutheusser:2021frk,Leutheusser:2021qhd} is that in this strict $G_N \to 0$ limit, one already sees the emergence of Killing horizons from the algebra of the quantized scalar field. Below, we push this picture further and explain how local geometric objects, including the spacetime metric and curvature tensor, emerge in this limit.

\subsection{Construction of a Length Function}
\label{kinematicsec}
Consider a minimally coupled non-interacting massive scalar field $\Phi$ of mass $m$ living on a $D=d+1$-dimensional curved spacetime $M$. Let $\mathcal{A}$ be the algebra of operators built out of the scalar field $\{\Phi(x)\}$ evaluated at points $x \in \mathcal{M}$, where $\Phi(x)$ is understood as an operator-valued distribution.\footnote{The technically precise objects are the smeared operators $\Phi(f) = \int \dd^Dx \sqrt{-g} \ \Phi(x) f(x)$, for $f \in C_0^\infty(\mathcal{M})$. Here, we work with pointwise $\Phi(x)$ and revert to smeared operators only where precision is required.} These operators act on a Fock space $\mathcal{H}$. Let $|\omega\rangle$ be a vector in this Hilbert space.  Motivated by the reconstruction program of Connes \cite{Connes1995}, we will work with the triple $(\mathcal{A},\mathcal{H}, |\omega\rangle)$ and show that this encodes \emph{all} the local information about the underlying geometry.

However, as we will see in the following, to get the geometry from these algebraic objects, one also needs to impose additional constraints on $|\omega\rangle$ and the operators in the algebra. In particular, we will assume that $|\omega\rangle$ satisfies the Hadamard condition. This requirement is very physical, as it is an operator algebraic realisation of the equivalence principle. Additionally, it can be shown that if we want the time derivative of the quantized fields to have finite fluctuations, then $|\omega\rangle$ must be Hadamard \cite{Fewster:2013lqa}. Moreover, a theorem of Wald and Kay states that whenever the background spacetime admits a Killing vector, one can always construct a Hadamard state on its maximal extension \cite{Kay:1988mu}.

In addition to this, we will also assume that the mass of the scalar field is known. In certain spacetimes, all the generators of the symmetries can be constructed from the operator algebraic data (see Appendix \ref{poincareappendix} for an example). In such cases, one can obtain the mass of the scalar field from the action of these generators on the scalar field through the standard relations. When the mass cannot be determined from the operator algebra, we simply assume that it is known.

With these algebraic inputs in hand, we now extract the relevant geometrical structures. The algebra $\mathcal{A}$ is spanned by a family of operators $\mathcal{O}_i$ obtained by evaluating the quantized scalar field at some point $x_i \in M$. We will treat $i$ as a continuous index and set aside its geometric origin. In this way, we regard each spacetime point as being associated with an operator in the algebra.

Once we have a notion of spacetime points in terms of these operators, we can introduce a notion of distance between them. Using the two-point functions $\langle \omega | \mathcal{O}_i \mathcal{O}_j | \omega \rangle$, we define a length function by taking two limits. First, we take the large mass limit, and generically find that 
\bea
\bra \omega | \mathcal{O}_i \mathcal{O}_j |\omega \ket_0 \sim e^{-m \ell(i,j)} \times  m^k\,,
\eea
where $k$ is some constant. The subscript on the left-hand side indicates that we are keeping only the leading order term in the large mass limit. Here $\ell$ is a distance function, which can then be extracted by taking the second limit:
\bea
\ell(i,j) = -\lim_{m \to 0} \left(\frac{1}{\bra \omega | \mathcal{O}_i \mathcal{O}_j |\omega \ket_0}\frac{\partial }{\partial m} \bra \omega | \mathcal{O}_i \mathcal{O}_j |\omega \ket_{0}\right)_{\text{reg.}} \,,\label{lengthexpressioneq}
\eea
We regularize the above expression by removing the $1/m$ divergent term before taking the limit. Now let us briefly explain the origin of \eqref{lengthexpressioneq}. The scalar field obeys the wave equation on the curved spacetime. The large mass limit corresponds to the geometrical optics limit in which the wave equation reduces to the geodesic equation (see Appendix A of \cite{Berenstein:2020vlp} for an explicit computation). Consequently, the two-point function is dominated by the contribution of a particle propagating along the geodesic connecting the insertion points, allowing us to extract the distance between them.\footnote{There is an important caveat here. When two points are sufficiently far apart, there need not exist a real valued geodesic connecting them. In such cases, the two-point function instead encodes the lengths of complex geodesics connecting the two points \cite{Fidkowski:2003nf}. This issue does not arise here, as we are only concerned with local geometric properties extracted from the coincidence limit of the distance function.}

If one were to start with some arbitrary triple $(\mathcal{A},\mathcal{H},|\omega\rangle)$, the length function $\ell$ can be used to diagnose whether the emergent structure associated with the operator algebra admits a genuine geometric interpretation. In particular, we expect the length function to satisfy the following properties
\bea
\ell(i,j) = \ell(j,i)\,, & \qquad \ell(i,i)=0\,.
\eea
In the case where $|\omega\rangle$ is a cyclic and separating vector with a geometric modular flow associated to it (see Appendix \ref{reviewapp} for definitions), we have more stringent constraints on the emergent geometry. Note that geometric modular flows are present when the action of the modular Hamiltonian $\hat{H}$ corresponds to a symmetry of the spacetime \cite{Buchholz1993,Buchholz:1998pv}. Let us define
\bea
\mathcal{O}_j = e^{-i \hat{H}s_1}\mathcal{O}_ie^{i \hat{H} s_1 }\,,\qquad \mathcal{O}_k = e^{-i \hat{H} s_2 }\mathcal{O}_ie^{i\hat{H} s_2 } \qquad \text{with} \ 0\leq s_2\leq s_1\,,
\eea
so that the corresponding spacetime points $x_{i,j,k}$ associated with the operator insertions are timelike separated. The emergent length function is then required to satisfy the reverse triangle inequality:
\bea
|\ell(i,j)| \geq & \ |\ell(i,k)| + |\ell(k,j)|\,.
\eea
This provides an additional constraint on the length functions associated to the operator algebras supported on a pseudo-Riemannian geometry. Also, if we keep $i,j$ fixed and search for all possible $\mathcal{O}_k$ that saturate the above inequality, then we can find the operators inserted along the timelike geodesic connecting $x_i$ and $x_j$. Therefore, the availability of a length function can be used to extract various important geometrical objects (see \cite{Rylov2002} for more details).

 Now, let us work out a few examples.

\noindent\textbf{Example 1} : Scalar field in Flat Space

The Wightman function of a free scalar field in $d+1$ Minkowski space can be explicitly solved for,
\bea
\bra \omega | \mathcal{O}_i \mathcal{O}_j |\omega \ket=\frac{(2 \pi)^{-\frac{d+1}{2}}}{\ell^{d-1}}\left(m \ell\right)^{\frac{d-1}{2}} K_{\frac{d-1}{2}}\left(m \ell\right)\,,
\eea
where $K_\nu$ is the modified Bessel function and $\ell$ is the flat-space geodesic distance. In the large mass limit, we have 
\bea
\bra \omega | \mathcal{O}_i \mathcal{O}_j |\omega \ket_0 \sim \frac{(2 \pi)^{-\frac{d}{2}}}{2 \ell^{d-1}}(m \ell)^{\frac{d-2}{2}} e^{-m \ell}
\eea
Plugging this expression into \eqref{lengthexpressioneq}, we can see that our definition reproduces the Minkowski length function.

\noindent\textbf{Example 2} : Scalar field in AdS$_{d+1}$

The Wightman function in the Euclidean space is given by \cite{DHoker:1998ecp}:
\bea
\bra \omega | \mathcal{O}_i \mathcal{O}_j |\omega \ket=2^{\Delta} C_{\Delta} \xi^{\Delta} F\left(\frac{\Delta}{2}, \frac{\Delta}{2}+\frac{1}{2} ; \Delta-\frac{d}{2}+1 ; \xi^2\right) .
\eea
where
\bea
\begin{aligned}
\xi = \frac{1}{\cosh \ell}\,, \qquad C_{\Delta}&=\frac{\Gamma(\Delta) \Gamma\left(\Delta-\frac{d}{2}-\frac{1}{2}\right)}{(4 \pi)^{(d+1) / 2} \Gamma(2 \Delta-d+1)}\,,\\
&m^2 = \Delta\left(\Delta-d\right)
\end{aligned}
\eea
Note that the characteristic AdS length scale is set to one. Taking the large mass limit by using an asymptotic expansion of the Hypergeometric function \cite{Paris:2013,Paris:2013ugv}, followed by Stirling's approximation, we obtain  
\bea 
\bra \omega | \mathcal{O}_i \mathcal{O}_j |\omega \ket_0 \sim m ^{\frac{d-4}{2}} e^{-m \ell}\,,
\eea
where we have dropped an $m$ independent overall constant. Therefore, \eqref{lengthexpressioneq} once again gives us the correct length function.

\subsection{Curvature from Local Flows}
\label{curvaturesection}

\begin{figure}
\centering
\includegraphics[width=0.95\linewidth]{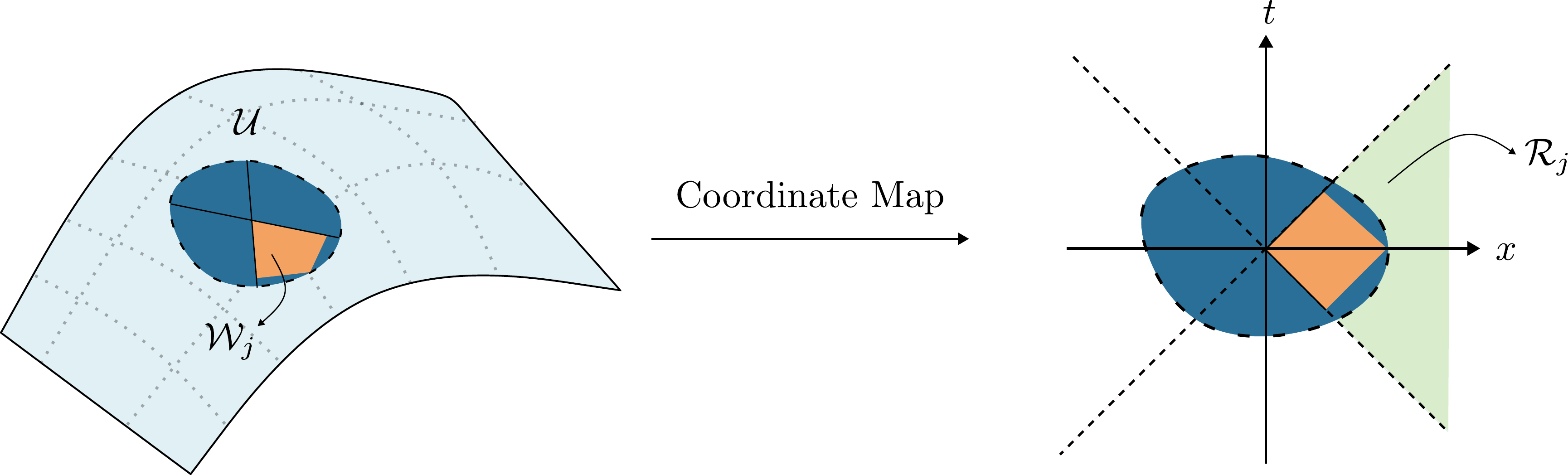}
\caption{\small Consider a locally flat patch $\mathcal{U}$ on the manifold. In this patch, we can identify local Rindler horizons that partition the patch into multiple partial Rindler wedges. We denote the algebra of operators restricted to a causal diamond in one such partial wedge by $\mathcal{W}_j$. On the right, we have the map of the patch to a globally flat spacetime. The partial Rindler wedge has a global extension in this Minkowski space. The algebra of operators of a massive scalar field on the full Rindler wedge is denoted by $\mathcal{R}_j$.}
\label{flatchartfig}
\end{figure}

In order to specify a metric-compatible (pseudo-) Riemannian geometry, one needs to have access to the metric and its derivatives at all points on the manifold. Once we have identified the length function on the manifold, it is straightforward to construct the metric tensor. First, let us introduce a quantity $\sigma$, often referred to as Synge's world function \cite{Synge:1960ueh},
\bea
\sigma(i,j) = \frac{\ell^2}{2}\,,
\eea
with $\ell(i,j)$ defined in \eqref{lengthexpressioneq}.
On a pseudo-Riemannian manifold the metric tensor can be obtained from the world function via 
\bea
\lim_{x\to x^{\prime}}\frac{\partial^2}{\partial x^{\mu} \partial x^{\prime}{}^{\nu} }\sigma(x,x^{\prime}) = g_{\mu \nu}\,,\label{metricfromsigmaeq}
\eea
where $x$ and $x^{\prime}$ are two points on the manifold \cite{DeWitt:1960fc}. Here, however, our ``points'' are merely labels $i,j,...$ indexing operators, with no a priori geometric meaning attached to them. Therefore, we can neither take derivatives nor coincident limits. To sidestep this difficulty, we use the equivalence principle. 
For every point $\mathcal{P}$ on the manifold, one can find a sufficiently small neighbourhood $\mathcal{U}$ in which a flat chart exists. Within this locally flat patch, we consider a family of Rindler wedges whose bifurcation surfaces all pass through the point $\mathcal{P}$. We denote by $\mathcal{W}_j$ the algebra of operators supported on these wedges. We will choose $\mathcal{W}_j$ to be supported in a causal diamond inside the Rindler wedge to make the algebra well-defined (see the left panel of Figure \ref{flatchartfig}). We then require these subalgebras to satisfy the following condition:
\begin{itemize}
\item
For each $\mathcal{W}_j$, there exists a pair
\bea
(\mathcal{R}_j,\omega_{\text{Mink}})\,,
\eea
where $\mathcal{R}_j$ is the operator algebra of a massive scalar field on an \emph{entire} $(d+1)$-dimensional Rindler spacetime, such that the correlation functions of operators in $\mathcal{W}_j$ in $|\omega\rangle$ are well approximated by the vacuum expectation values of local operators in $\mathcal{R}_j$. In this sense, every operator in $\mathcal{W}_j$ can be approximated by a local operator in $\mathcal{R}_j$.
\end{itemize}

\noindent
We will refer to the subalgebras $\mathcal{W}_j$ as \emph{local Rindler algebras}. There is a natural geometrical way to understand each $\mathcal{W}_j$ and its corresponding Rindler algebra $\mathcal{R}_j$ (see Figure \ref{flatchartfig}). The locally flat patch $\mathcal{U}$ can be divided into various Rindler wedges, and each $\mathcal{W}_j$ corresponds to the algebra of operators living in these wedges. The flat coordinates we introduce on $\mathcal{U}$ admit a global Minkowski extension, and the Rindler wedges in $\mathcal{U}$ can similarly be extended to full Rindler spacetimes using the same coordinate chart. The algebras $\mathcal{R}_j$ are then precisely the algebras of operators on these extended Rindler spacetimes. Note that, by construction, $\mathcal{R}_j$ is a much larger set than $\mathcal{W}_j$. 

If one has access to algebras of operators in multiple Rindler wedges of a globally flat Minkowski spacetime, then the entire Poincar\'{e} algebra of the spacetime can be reconstructed purely from the operator algebras \cite{Borchers1996,Brunetti:1992zf}. We briefly outline the steps here but an explicit example is worked out in Appendix~\ref{poincareappendix}. By the Bisognano--Wichmann theorem \cite{Bisognano:1975ih,Bisognano:1976za}, one can associate a modular operator $\Delta_j$ to each wedge algebra, and these operators generate boosts that preserve the respective wedges. The boost generators of the local Poincar\'{e} algebra are then given by these modular operators. For each wedge, one can further construct subalgebras with an inclusion structure, which leads to half-sided modular translations. These can, in turn, be used to construct local translation generators. Combining the boost operators associated with different wedges with these translation generators, one can reconstruct rotations and, ultimately, the full Poincar\'{e} group.

This construction allows us to use the $\mathcal{R}_j$ algebras to construct the Poincar\'{e} algebra associated to the global Minkowski extension of the patch $\mathcal{U}$. Since we assume the operators in $\mathcal{W}_j$ to be approximately equal to local operators in $\mathcal{R}_j$, these emergent generators act as approximate symmetry generators on $\mathcal{W}_j$. Therefore, the assumption of local Rindler algebras implies that matter effectively ``sees'' flat space locally, thereby realizing the equivalence principle via operator algebras.

But most importantly, the emergence of these approximate symmetry generators allows us to identify approximate Killing flows in the local patch $\mathcal{U}$ purely in the language of operator algebras. Using these flows, we can then define derivatives and a coincident limit. As an example, consider the boost generated by one of the modular operators $\Delta_j$, and let $\xi^{\mu}$ denote the corresponding approximate Killing vector field.

Now consider an operator $a_{\text{Rind}} \in \mathcal{R}_j$. Its modular evolution is given by
\bea
a_{\text{Rind}}(s) = \Delta_j^{is} \ a_{\text{Rind}} \ \Delta_j^{-is}\,.
\eea
Since $a_{\text{Rind}}(s)$ is the image under the modular flow of an \emph{exact} Rindler spacetime, the evolution corresponds to a geometric flow along the vector field $\xi^{\mu}$, and hence defines a one-parameter family of operators associated with points along this flow. The existence of a map from the operators in $\mathcal{W}_j$ to local operators in $\mathcal{R}_j$ and vice versa automatically induces a one-parameter family of operators $\mathcal{O}(s)$ for every $\mathcal{O} \in \mathcal{W}_j$. The important point to note here is that $\mathcal{O}(s)$ for every value of $s$ is a local operator in $\mathcal{W}_j$.

Therefore, the existence of a map to a true Rindler algebra allows us to associate a pointwise flow within $\mathcal{W}_j$. The continuous parameter $s$ labeling this flow then enables us to define derivatives and, in particular, a coincident limit. Consider the following two-point function
\bea
\bra \omega | \mathcal{O}_i(s_1)\mathcal{O}_i(s_2) |\omega \ket\equiv G(i_{s_1},i_{s_2})\,.
\eea
Then by following the procedure in Section \ref{kinematicsec}, we can compute $\sigma(i_{s_1},i_{s_2})$. Since $s_{1,2}$ are continuous parameters, we find that
\bea
\lim_{s_1,s_2\to 0}\frac{\partial^2}{\partial s_1 \partial s_2 }\sigma(i_{s_1},i_{s_2}) = g_{\mu \nu} \xi^{\mu} \xi^{\nu}\,. \label{emergentmetriceq}
\eea
Therefore, we obtain the contraction of the metric with respect to one of the approximate Killing vectors of the locally flat patch. Repeating the same computation for the other local flows gives the contractions of the metric with respect to the remaining approximate Killing vectors. Since this can be done for \emph{all} the approximate symmetries of the local patch, we have sufficient information to determine all components of the metric tensor.

Now, let us compute curvature tensors from the operator algebra. The idea is to use the fact that curvature tensors appear in the subleading terms of the coincidence limit of the Wightman function \cite{Decanini:2005eg}. To see this, let us consider the two-point function
\bea
\bra \omega | \mathcal{O}_i \mathcal{O}_i(s) |\omega \ket \equiv G(i,i_s)\,, \label{modularevolvedtwopt}
\eea
where the operator $\mathcal{O}_i(s)$ is the family of operators corresponding to an operator insertion on the curve generated by the action of a local Rindler boost. We can see that taking $s \to 0$ then gives us the coincidence limit of the two-point function.

In this limit, the two-point functions have the following universal Hadamard form \cite{Decanini:2005eg},
\bea
\bra \omega | \mathcal{O}_i \mathcal{O}_{j} |\omega \ket=\frac{i \alpha}{2}\left(\frac{U\left(i,j\right)}{\left[\sigma\left(i,j\right)+i \epsilon\right]^{D / 2-1}}+V\left(i,j\right) \ln \left[\sigma\left(i,j\right)+i \epsilon\right]+W\left(i,j\right)\right) ,\label{Hadamardformeq}
\eea
where $\alpha$ is a constant that depends only on the dimensions of the spacetime. Note that the logarithmically divergent piece in \eqref{Hadamardformeq} shows up only in even spacetime dimensions. Now let us replace $\mathcal{O}_j$ with the operator evolved by the local Rindler boost for some modular time $s$, as in \eqref{modularevolvedtwopt}.

In the $s \to 0$ limit, the coefficient $U$ goes to $1$ and the leading behaviour approaches the propagator of a massless scalar field in Minkowski space:
\bea
G(i,i_s) = \frac{i\alpha}{2}\frac{1}{\left[\sigma\left(i,i_s\right)+i \epsilon\right]^{D / 2-1}}\,.
\eea
In particular, the number of spacetime dimensions $D$ appears in the leading term. 

Now let us look at the subleading behaviour of the two-point function. It turns out that we can expand the function $U$ in powers of $\sigma$ \cite{Decanini:2005eg},
\bea
U(i,j) = \Delta_{V}^{1/2} +O(\sigma)\,,
\eea
where $\Delta_V$ is the van Vleck-Morette determinant, and this quantity in turn has the following expansion \cite{DeWitt:1960fc},
\bea
\Delta_V \approx 1+ \frac{1}{6}R_{\mu \nu} \sigma^{\mu} \sigma^{\nu}\,, \qquad \text{where} \quad \sigma^{\mu} \approx s \xi^{\mu}\,.
\eea
Therefore, we can extract the Ricci tensor from the two-point functions as follows,
\bea
\lim_{s\to 0} \frac{\partial^2}{\partial s^2}\frac{12\sigma^{D/2-1}}{i\alpha}G(i,i_s) = R_{\mu \nu} \xi^{\mu} \xi^{\nu}\,. \label{emergentriccieq}
\eea

\noindent
There is a simple reason why the Ricci tensor can be extracted from the two-point functions of the theory. The van Vleck-Morette determinant arises from the one-loop correction to the semiclassical approximation of the propagator, which contains information about the focusing/defocusing of geodesics. This, in turn, means it contains information about the Ricci tensor through the Raychaudhuri equation \cite{Visser:1992pz}.

Let us recap our results, so far. Under the conditions \hyperref[item:first]{(1)}-\hyperref[item:third]{(3)}, the metric and curvature tensors at every point on the manifold can be extracted entirely from algebraic data. The derivation assumed the existence of an underlying geometry, in particular, a local Rindler horizon structure. Yet the conditions \hyperref[item:first]{(1)}-\hyperref[item:third]{(3)} themselves are formulated purely in the language of operator algebras and their associated Hilbert spaces. The absence of any geometrical input in these conditions means that they can be read equally as criteria for the \emph{emergence of spacetime}. One may begin with an abstract triple $(\mathcal{A}, \mathcal{H}, |\omega\rangle)$ carrying no \emph{a priori} geometric interpretation and ask whether conditions \hyperref[item:first]{(1)}-\hyperref[item:third]{(3)} are satisfied. If they are, a spacetime geometry can be systematically constructed from the algebraic data. 

\subsection{Einstein's Equations from Locally Stationary States}
\label{dynamicsec}

In the previous subsection, we identified the conditions under which a spacetime geometry emerges from the operator algebra and Hilbert space of a quantized massive scalar field. The resulting geometry need not satisfy Einstein's field equations. As we are in the strict $G_N \to 0$ limit, there is, however, a simple check for the validity of Einstein's equations. In this limit, they reduce to the vacuum equations
\bea
R_{\mu \nu} = \frac{2\Lambda}{D-2} g_{\mu \nu}\,.
\eea 
Using the expressions for the emergent metric \eqref{emergentmetriceq} and Ricci tensor \eqref{emergentriccieq}, the contraction of Einstein's equations with the approximate Killing vectors of a locally flat patch can be written as
\bea
\lim_{s \to 0}\frac{\partial^2}{\partial s^2}\frac{24\sigma^{D/2-1}}{i\alpha}G(i,i_s)= \left(\frac{2\Lambda}{D-2} \right)\lim_{s_1,s_2\to 0}\frac{\partial^2}{\partial s_1 \partial s_2 }\sigma(i_{s_1},i_{s_2})\,.
\eea
Choosing different approximate Killing vectors results in a set of equations that, when satisfied simultaneously, 
guarantee the validity of the vacuum Einstein equations.

We have arrived at an operator-algebraic diagnostic for Einstein's field equations, but why should they emerge in the first place?  We now show that an additional operator-algebraic condition is sufficient to derive them, provided we include perturbative $G_N$ corrections around the $G_N \to 0$ limit so that the scalar field back-reacts on the geometry (see \cite{Kudler-Flam:2023qfl,Jensen:2023yxy,Wall:2011hj,Klinger:2026tws} for more details on this limit). 

\begin{figure}
\centering
\includegraphics[width=0.45\linewidth]{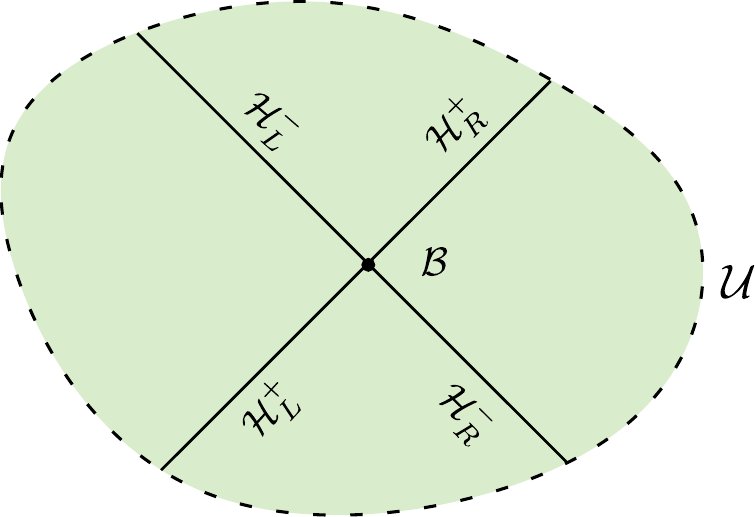}
\caption{\small Consider a Rindler horizon in a locally flat patch $\mathcal{U}$ on the manifold. We can divide the horizon into four null hyperplanes as shown in the figure. The bifurcation surface is denoted by $\mathcal{B}$.}
\label{horizonsfig}
\end{figure}

A major breakthrough in understanding the origin of Einstein's field equations came in the seminal work of Jacobson \cite{Jacobson:1995ab}. We can always find a small neighbourhood of a spacetime point $\mathcal{P}$ where a locally flat chart exists and consider an accelerating observer in this patch. In any such local Rindler frame, Einstein's equations can be derived from a local equilibrium condition via the Clausius relation \cite{Jacobson:1995ab},
\bea
\delta Q = T \dd S\,, \label{Clausiusrelation}
\eea
where $\delta Q$ is the heat flux across the local Rindler horizon, $T$ is the corresponding Unruh temperature, and $\dd S$ is the change in the horizon entropy. Jacobson's argument clarifies the emergent nature of gravity: gravity is a theory of local Rindler horizons and their associated thermodynamics (see also \cite{Padmanabhan:2008zzc}). 

Our goal, is to identify an operator-algebraic analogue of Jacobson's Clausius relation \eqref{Clausiusrelation}. We start with the Weyl algebra $\mathcal{W}$ generated by bounded operators $W(f) = e^{i\Phi(f)}$, where $f \in C_0^\infty(\mathcal{M})$. Now consider a locally flat patch $\mathcal{U}$ in the emergent geometry. As in the previous section, we find a local Rindler horizon in this patch composed of two null hypersurfaces $\mathcal{H}^{\pm}$ and a bifurcation surface $\mathcal{B}$. Each hypersurface decomposes as the union of a left and right hyperplane as $\mathcal{H}^{\pm} = \mathcal{H}^{\pm}_L\cup\mathcal{H}^{\pm}_R$ (see Figure \ref{horizonsfig}). Consider the von Neumann algebra $\mathcal{M_R}$ associated to the right Rindler wedge of this local patch \cite{Dorau:2025hmq}. Let $\mathcal{N}_R \subset \mathcal{M}_R$ be the von Neumann algebra of operators restricted to $\mathcal{H}^{-}_R$. 

Consider a state $\omega_0$ on $\mathcal{W}$ (see Appendix \ref{reviewapp} for the definition of a state). We now assume that $\omega_0$ is a quasifree\footnote{A quasifree, or Gaussian, state is one in which all odd $n$-point functions vanish, while the even $n$-point functions are completely determined by Wick contractions (see \cite{Khavkine:2014mta,Kay:1988mu} for further details).} Hadamard state stationary under the affine dilations along $\mathcal{H}^{-}_R$. We refer to such states as \emph{locally stationary states}.

The restriction of the state $\omega_0$ to the horizon algebra $\mathcal{N}_R$, which we denote by $\left.\omega_{0}\right|_{\mathcal{N}_R}$, satisfies the Kubo-Martin-Schwinger (KMS) condition at an inverse temperature $\beta$:
\bea
\left.\omega_{0}\right|_{\mathcal{N}_R}(a_1 \alpha_{t+i\beta}[a_2]) = \left.\omega_{0}\right|_{\mathcal{N}_R}(\alpha_{t}[a_2]a_1)\,, \qquad a_{1,2} \in \mathcal{N}_R \ \text{and} \ t \in \mathbb{R}\,,
\eea
where $\alpha_t[a]$ is the flow of $a \in \mathcal{N}_R$ under affine dilations along $\mathcal{H}^{-}_R$ \cite{Summers:1995kp}. The inverse temperature $\beta$ is given by $\frac{2\pi}{\kappa}$, where $\kappa$ is the surface gravity of $\mathcal{H}^{-}_R$. Now consider the Gelfand--Naimark--Segal (GNS) representation of $(\mathcal{N}_R,\left.\omega_{0}\right|_{\mathcal{N}_R})$. We denote the vector associated to $\left.\omega_{0}\right|_{\mathcal{N}_R}$ by $|\Omega_0\rangle$. The KMS condition implies that the GNS vector $|\Omega_0\rangle$ is separating for $\mathcal{N}_R$, so that Tomita-Takesaki theory applies and the modular operator $\Delta_{\Omega_0}$ is well-defined. By the Bisognano-Wichmann theorem as extended to Killing horizons \cite{Summers:1995kp,Dorau:2025hmq}, $\Delta_{\Omega_0}$ acts as affine dilations along $\mathcal{H}^{-}_R$.
Finally, let $\omega_\phi$ be a coherent excitation of $\omega_0$. In the GNS Hilbert space, this state is represented by $|\Omega_{\phi}\rangle = e^{i\Phi(\phi)}|\Omega_0\rangle$, where $\phi$ is a smooth solution of the Klein-Gordon equation with support only on $\mathcal{H}^{-}_R$. 

We now include perturbative $G_N$ corrections around the $G_N \to 0$ limit. In this perturbative quantum gravity regime, observables are not fully diffeomorphism invariant, but are instead invariant under the symmetries of the fixed background geometry. In our case, we require observables to be invariant under the flow generated by $\Delta_{\Omega_0}$. To construct such observables, one must appropriately ``dress'' the operators with the charges associated with this flow (see \cite{Jensen:2023yxy,Kudler-Flam:2023qfl,Klinger:2026tws} for details). We will not review this construction here, but note that the resulting algebra is a crossed product algebra (see Section \ref{typeiconstructionsection} for definitions), and is of type II. This allows us to define density matrices, traces, and entropy for states in the algebra.

In this crossed product algebra, we can then define a state $|\hat{\Omega}_\phi\rangle$ corresponding to the coherent excitation $|\Omega_\phi\rangle$ as follows:
\bea
|\hat{\Omega}_\phi\rangle=\int_{\mathbb{R}} \dd x \ f(x)\left|\Omega_\Phi\right\rangle \otimes|x\rangle\,,
\eea
where $f(x) \in L^{2}(\mathbb{R})$ and $|x\rangle$ denotes the generalized position basis of $L^{2}(\mathbb{R})$. We make this state \emph{semiclassical} by choosing the function $f(x)$ to vary `slowly' in $x$. Then it follows that
\bea
S^{\mathrm{rel}}\left(\omega_{0} \| \omega_{\phi}\right) = \left\langle\frac{A}{4 G_{\mathrm{N}}}\right\rangle_{\hat{\Omega}_{\phi}} -\frac{A}{4 G_{\mathrm{N}}}+ S_{\mathrm{vN}}\left(\left.\omega_\phi\right|_{\mathcal{H}_{\mathrm{R}}^{-}}\right)-S_{\mathrm{vN}}\left(\left.\omega_0\right|_{\mathcal{H}_{\mathrm{R}}^{-}}\right)\,, \label{mainjacobsoneq}
\eea
where $A$ is the `background area' of the bifurcation surface \cite{Kudler-Flam:2023qfl}. The quantity on the left-hand side is the relative entropy between the two states. The von Neumann entropy on the right-hand side corresponds to the entanglement entropy of the states restricted to the past of the bifurcation surface. A version of this equality was originally used in proving the generalized second law for causal horizons \cite{Wall:2011hj} (see \cite{Jensen:2023yxy} for the derivation of this expression in the case of operators restricted to a causal diamond).

There are two important technical points that we would like to emphasize before proceeding. First, the relative entropy on the left-hand side of \eqref{mainjacobsoneq} is well defined, whereas the individual terms appearing on the right-hand side are not separately well defined in the $G_N \to 0$ limit. The claim, however, is that the particular combination of these terms remains well defined even in the $G_N \to 0$ limit \cite{Susskind:1994sm}. Second, our definition for the expectation value of the area operator in \eqref{mainjacobsoneq} differs from \cite{Kudler-Flam:2023qfl}. There, it includes the background area $A$, along with a perturbative correction from the inclusion of the charge associated with $\Delta_{\Omega_0}$, as well as contributions from the matter flux through the horizon. However, since the perturbative charge correction drops out of the computation of the relative entropy, we omit this term in the definition of the expectation value of the area for brevity.

Now let us look at the left-hand side of \eqref{mainjacobsoneq}. For coherent excitations, the relative entropy is given by
\bea
S^{\mathrm{rel}}\left(\omega_0 \| \omega_\phi\right)=\left.i \frac{d}{d t}\right|_{t=0}\left\langle\Omega_\phi \mid \Delta_{\Omega_0}^{i t} \Omega_\phi\right\rangle .
\eea
In \cite{Kurpicz:2021kgf} it was shown that this expression can be rewritten as 
\bea
S^{\mathrm{rel}}\left(\omega_0 \| \omega_\phi\right)=-2 \pi \int_{\mathcal{H}^{-}_R} U\left\langle: T_{\mu \nu}:\right\rangle k^\mu k^\nu \dd U \dd \mathrm{vol}_{\mathcal{S}}\,, \label{deltaqexp}
\eea
where $k^{\mu}$ is the tangent vector to the horizon, $\langle:T_{\mu\nu}:\rangle$ is the expectation value of the stress energy tensor of the quantum field in the state $\omega_{\phi}$ and $U$ corresponds to the null coordinate along the horizon. The energy flux through the Rindler horizon is then given by \cite{Dorau:2025hmq,Kurpicz:2021kgf}
\bea
\delta Q = \frac{S^{\mathrm{rel}}\left(\omega_0 \| \omega_\phi\right)}{\beta}\,.\label{heatfluxexp}
\eea 
In order to determine the right-hand side of \eqref{mainjacobsoneq}, we observe that for coherent excitations of the Minkowski vacuum, the entanglement entropy remains unchanged \cite{Fiola:1994ir,Benedict:1995yp}. Therefore the difference in the von Neumann entropy of the states drops out, and we are left with
\bea
S^{\mathrm{rel}}\left(\omega_{0} \| \omega_{\phi}\right) =\frac{\delta A}{4\pi G_N}\,. \label{deltaSexp}
\eea
Using \eqref{heatfluxexp} and \eqref{deltaSexp}, we obtain Jacobson's Clausius relation \eqref{Clausiusrelation}.

We complete the derivation of Einstein's equation by noting that 
\bea
\frac{\delta A}{4\pi G_N}=\frac{1}{4G_N} \int_{\mathcal{H}^{-}_R}\theta \dd U \dd \mathrm{vol}_{\mathcal{S}}\,,
\eea
where $\theta$ is the expansion parameter of the null congruence that forms the local Rindler horizon. This gives us
\bea
\frac{1}{4G_N} \int_{\mathcal{H}^{-}_R}\theta \dd U \dd \mathrm{vol}_{\mathcal{S}} =-2 \pi \int_{\mathcal{H}^{-}_R} U\left\langle :T_{\mu \nu}:\right\rangle k^\mu k^\nu \dd U \dd \mathrm{vol}_{\mathcal{S}}\,.
\eea
Imposing conservation of the stress energy tensor and using the Raychaudhuri equation, we arrive at the  Einstein's equation in standard form \cite{Jacobson:1995ab},
\bea
R_{\mu \nu}-\frac{1}{2} R g_{\mu \nu}+\Lambda g_{\mu \nu}=8\pi G_N \left\langle :T_{\mu \nu}:\right\rangle\,.
\eea

\section{Type I Algebras from Random Matrix Theory Corrections}
\label{typeIsection}
In the previous section, we worked in the $G_N \to 0$ limit and incorporated perturbative corrections in $G_N$ round this limit. When we instead turn on a small but finite $G_N$, non perturbative quantum gravity corrections become important. In this section, we show how these corrections can be systematically incorporated by using holography.

\subsection{Microstates from a Boundary Dual}
\label{boundarymicrostatesec}

In recent years, there have been several constructions of microstates of non-supersymmetric black holes \cite{Balasubramanian:2022lnw,Balasubramanian:2022gmo,Climent:2024trz,Banerjee:2024fmh,Magan:2024nkr,Geng:2024jmm}. In this subsection, we briefly review the considerations of \cite{Banerjee:2024fmh,Magan:2024nkr}, where non-perturbative corrections, obtained via random matrix theory, are used to arrive at a finite-dimensional Hilbert space of black hole microstates.

Consider a pair of CFTs in the thermofield double (TFD) state dual to an eternal AdS black hole,
\bea
\left|\text{TFD} \right\rangle = \frac{1}{\sqrt{Z(\beta)}}\sum_n e^{-\frac{\beta}{2} E_n} |E_n\rangle_{L} |E_n\rangle_{R}\,,\label{canonicaltfdstateeq}
\eea
where $|E_n\rangle_{L,R}$ are the energy eigenstates of the left and the right CFT, and the partition function is given by
\bea
Z(\beta) = \sum_n e^{-\beta E_n}\,.
\eea
We can construct an infinite family of time-shifted TFD states by acting on the TFD state only with the right CFT Hamiltonian,
\bea
\left|\text{TFD}(t) \right\rangle =e^{-i H_R t}\left|\text{TFD} \right\rangle = \frac{1}{\sqrt{Z(\beta)}}\sum_n e^{-\frac{\beta}{2} E_n} e^{-i t E_n} |E_n\rangle_{L} |E_n\rangle_{R}\,.
\eea
These states are closely related to the thin-shell states studied in \cite{Balasubramanian:2022gmo,Climent:2024trz}, where instead of acting with the right boundary Hamiltonian, one inserts a heavy operator. Time-evolved TFD states were also used in the state-dependent reconstruction of interior operators of an eternal black hole in \cite{Papadodimas:2015xma}.
Now consider the Hilbert space spanned by a discrete set of $n$ time-shifted TFD states,
\bea
\mathcal{H}_n = \text{span}\{|\text{TFD}(t_j)\rangle : t_j = j t_1 \ \text{and} \ j=1,\dots n\}\,, \label{nhilbertspace}
\eea 
for a fixed time-step $t_1$ taken sufficiently large (a few times the inverse temperature $\beta$) such that all $|t_i - t_j|$ are well-separated \cite{Magan:2024nkr}.  To determine the dimension of this Hilbert space, we study the Gram matrix,
\bea
G_{ij} =\left\langle \text{TFD}(t_i)|\text{TFD}(t_j)\right\rangle \,.
\eea
The dimension of $\mathcal{H}_n$ is then given by 
\bea
d_{n} = n-\ker(G)\,,
\eea
where $\ker(G)$ is the dimension of the kernel of the Gram matrix. In the $N \to \infty$ limit of the boundary CFT, $G_{ij} = \delta_{ij}$ \cite{Banerjee:2024fmh}. As a result, $d_n = n$, and the dimension diverges in the limit $n \to \infty$.

At finite $N$, however, the overlaps receive non-perturbative corrections and in principle one needs control over the boundary theory at finite $N$ to describe the physics. As it turns out, finite $N$ effects can be modelled within the $N\to \infty$ limit itself by treating the large $N$ theory as a random matrix theory (RMT) \cite{Magan:2024nkr,Banerjee:2024fmh}. We implement this approximation in two steps. We first assume that in the $N \to \infty$ limit, the CFT admits a continuous density of states $\rho_0(E)$. Operationally, this amounts to making the following replacement:
\bea
\sum_{n} \ \to \ \int \dd{E}  \ \rho_0(E)\,, \qquad \ \text{and} \qquad \ |E_n\rangle \ \to \ |E\rangle\,. \label{replacementeq}
\eea
The density of states can be computed from the partition function through an inverse Laplace transform,
\bea
\rho_{0}(E) = \lim_{N\to\infty}\frac{1}{2\pi i}\int^{\gamma+i\infty}_{\gamma-i\infty} \dd{\beta}\ e^{\beta E} Z(\beta)\,. \label{canonicaldoseq}
\eea
We then substitute the density of states $\rho_0(E)$ with the density of states $\rho(E)$ of a random matrix theory, satisfying $\overline{\rho(E)} = \rho_0(E)$. For now, we choose the higher point functions of the RMT to follow a Poisson distribution \cite{Magan:2024nkr},
\bea
\overline{\rho(E) \rho\left(E^{\prime}\right)}=\rho_0(E) \rho_0\left(E^{\prime}\right)+ \rho_{0}(E) \delta\left(E-E^{\prime}\right)\,.\label{2ptdensityofstateseq}
\eea
We refer to the inclusion of RMT corrections via this procedure as an \emph{RMT completion} of the theory. An important consequence of the completion is the modification of the overlap of the time-shifted TFD states. Consider
\bea
\left|G_{ij}\right|^2=\left|\left\langle \text{TFD}(t_i)|\text{TFD}(t_j)\right\rangle \right|^2= \frac{1}{Z(\beta)^2}\sum_{n,m} e^{-\beta(E_n+E_m)} e^{-i (t_j-t_i)(E_m-E_n)}\,.\label{overlapsquaredeq}
\eea
The expression on the right-hand side is called the spectral form factor (SFF). Making the replacement \eqref{replacementeq} and averaging, we get
\bea
\overline{\left|G_{ij}\right|^2} = \frac{\left|Z(\beta-i (t_i-t_j))\right|^2}{Z(\beta)^2}+\frac{Z(2\beta)}{Z(\beta)^2}\,.\label{secondoverlapeq}
\eea
The first disconnected piece decays for large $(t_i-t_j)$ and produces the `dip' in the SFF. The second term is constant and gives rise to the `plateau' in the SFF (see \cite{Cotler:2016fpe,Saad:2019pqd} for more details). The second term can be evaluated in a saddle point approximation, using
\bea
Z(\beta) = e^{-\beta M+S_{\text{BH}}}\,, \label{partitionfuntionsaddlepointeq}
\eea
where $M$ is the mass of the black hole. This gives us
\bea
\left.\overline{\left|G_{ij}\right|^2}\right|_c = e^{-S_{\text{BH}}} = e^{-O(N^2)}\,.\label{exponentialsuppoverlap}
\eea
The rank of $G$ can be computed from the connected component of these overlaps using resolvent techniques and one finds that the dimension of the RMT-completed Hilbert space $\overline{\mathcal{H}_n}$ saturates at $d_n =\min(n,e^{S_{\text{BH}}})$ \cite{Penington:2019kki,Balasubramanian:2022gmo,Banerjee:2024fmh}.
Now define 
\bea
\mathcal{H}_{\text{BH}} = \lim_{n\to\infty} \overline{\mathcal{H}_n}\,. \label{blackholehilbertspace}
\eea
The dimension of $\mathcal{H}_{\text{BH}}$ is then given by $d=e^{S_{\text{BH}}}$. Therefore, RMT completion results in a finite-dimensional Hilbert space carrying precisely the Bekenstein-Hawking entropy. 

We have arrived at a finite-dimensional Hilbert space by including RMT corrections. In the $N\to\infty$ limit, information about the discreteness of the CFT energy spectrum is lost, since level spacings are suppressed as $e^{-O(N^2)}$. One is therefore forced to work with a smooth density of states $\rho_0(E)$, and it is this `information loss' that leads to the orthogonality of the time-shifted TFD states.\footnote{The continuum spectrum is also at the root of Maldacena's large $N$ information problem \cite{Maldacena:2001kr}.} At any finite $N$, however, the spectrum is discrete, and the spectral form factor necessarily stops decaying and saturates \cite{Cotler:2016fpe,Saad:2019pqd}. One role of the RMT completion is to restore the discreteness of the spectrum at finite $N$. 

One can alternatively motivate the introduction of an RMT description in the boundary theory by appealing to \emph{chaos} and \emph{thermalization}. From the bulk perspective, this is quite natural, since black holes are known to be maximally chaotic systems in the sense of fast scrambling of quantum information \cite{Sekino:2008he,Shenker:2013pqa,Maldacena:2015waa,Cotler:2016fpe,Lowe:2024quv}. On the CFT side, the use of RMT dynamics has gained prominence in recent years \cite{Collier:2019weq,Cotler:2020ugk,Belin:2020hea,Haehl:2023tkr,DiUbaldo:2023qli,Collier:2023cyw,Jafferis:2024jkb}.
The emergence of an RMT description in the semiclassical limit need not be tied specifically to gravity or black holes. Rather, RMT provides a general framework for modeling chaotic dynamics in quantum systems. A classic example is a gas of hard spheres in a box, which is an integrable quantum system but nevertheless exhibits emergent RMT behaviour in the semiclassical limit \cite{Srednicki:1994mfb,Krishnan:2021faa}. From this perspective, the discussion in this and the following sub-sections extends to a broader class of chaotic systems.

With $\mathcal{H}_\text{BH}$ in hand, we turn to the algebra $\mathcal{B}(\mathcal{H}_\text{BH})$ of bounded operators acting on it. The finite dimensionality of $\mathcal{H}_\text{BH}$ strongly suggests this algebra is of type I, as argued in \cite{Banerjee:2024fmh}, via the discreteness of the spectrum. In the next subsection, we make this more explicit by constructing the minimal projectors required for a type I algebra.

\subsection{Crossed Product and the Construction of a type \texorpdfstring{I$_d$}{Id} Algebra}
\label{typeiconstructionsection}

The microstates of the previous section were generated by acting with $H_R$ on the TFD state, so a natural next step is to enlarge the large $N$ operator algebra by adjoining $H_R$ to it. In the large $N$ limit of $\mathcal{N}=4$ supersymmetric Yang-Mills theory, the relevant starting point is the algebra of single-trace operators and adjoining $H_R$ to it is made precise by the crossed product construction \cite{Witten:2021unn}. 

Consider the boundary CFTs in the previous section and pick an energy eigenstate $|E_0\rangle$ with energy $E_0 = O(N^2)$. Then a microcanonical version of the TFD state, centered around $|E_0\rangle$, can be defined as follows \cite{Chandrasekaran:2022eqq},
\bea
|\widetilde{\mathrm{TFD}}\rangle=e^{-S_{\text{BH}}/2} \sum_i e^{-\beta\left(E_i-E_0\right) / 2} f\left(E_i-E_0\right)\left|E_i\right\rangle_L\left|E_i\right\rangle_R\,. \label{microtfdeq}
\eea
Here, $f$ is a smooth, square-integrable function that can be chosen to be a Gaussian function with standard deviation $\sigma =O(N^0)$ without loss of generality.
We denote by $\mathcal{A}_{0,R}$ the algebra of single-trace operators in the right CFT. Working above the Hawking-Page temperature, this algebra becomes type III$_1$ in the $N\to \infty$ limit \cite{Leutheusser:2021frk,Leutheusser:2021qhd,Furuya:2023fei}. We then consider the renormalized right CFT Hamiltonian,
\bea
h_R = H_R-E_0\,.
\eea
The action of $h_R$ on the microcanonical TFD state produces fluctuations that are $O(N^0)$ \cite{Chandrasekaran:2022eqq}. If one were using the standard canonical TFD state, the fluctuations of the right CFT Hamiltonian would diverge in the large $N$ limit. Therefore, working with the microcanonical TFD state is advantageous as we can directly add $h_R$ to the single-trace algebra even in the strict $N \to \infty$ limit. We can \emph{formally} decompose the right Hamiltonian as follows
\bea
h_R = \hat{h} + X\,, \qquad\text{with}\  X = h_L \equiv H_L-E_0\,,
\eea
where $H_L$ is the Hamiltonian of the left CFT and the operator $\hat{h}=H_R-H_L$ is the modular Hamiltonian of the TFD state (rescaled by a factor of $1/\beta$). The extended Hilbert space takes the form $\mathcal{H} \cong \mathcal{H}_{\mathrm{TFD}} \otimes L^2(\mathbb{R})$,
where $\mathcal{H}_{\mathrm{TFD}}$ is the GNS Hilbert space generated by the action of single-trace operators on the TFD state. The appearance of this tensor product structure follows from the fact that the renormalized left Hamiltonian $h_L$ commutes with all operators in the right CFT. Consequently, its action furnishes an independent tensor factor.
The algebra acting on the extended Hilbert space is known in the literature as the crossed product algebra,
\bea
\mathcal{A}_R = \mathcal{A}_{0,R} \rtimes \mathbb{R}_h\,.\label{crossedprodcutalgebraeq}
\eea
It is generated by elements of the form $a \otimes 1$ with $a \in \mathcal{A}_{0,R}$, together with unitary operators of the form $e^{i \hat{h} s} \otimes e^{i X s}$. As a result, a general element of the algebra can be written as 
\bea
\hat{a} = \int_{-\infty}^{\infty} \dd{s} \  a(s) e^{is(\hat{h}+X)}\,, \qquad a(s) \in \mathcal{A}_{0,R} \,,
\eea
and a general state in the extended Hilbert space $\mathcal{H}$ has the form 
\bea
|\hat{\Phi}\rangle = \int_{-\infty}^{\infty} \dd{x} \ g(x) |\Phi\rangle|x\rangle \,, \qquad |\Phi\rangle \in \mathcal{H}_\text{TFD}\,.\label{generalstatecrossedprodcuteq}
\eea
Here $g(x) \in L^2({\mathbb{R}})$ is a normalizable function and $|x\rangle$ are states on which the operator $X$ acts as a position operator. The microcanonical TFD state is then identified with 
\bea
|\hat{\Psi}\rangle = \int_{-\infty}^{\infty} \dd{x} \ f(x) |\Psi\rangle|x\rangle \,,\label{tfdcrossedproducteq}
\eea
where $|\Psi\rangle$ is some element of the single-trace Hilbert space $\mathcal{H}_{\text{TFD}}$.

The crucial point here is that the resulting enlarged algebra is a type II$_{\infty}$ factor, and as a result, we can define density matrices and a trace on it. In particular, for \emph{any} $|\hat{\Phi}\rangle \in \mathcal{H}$, we can define a density matrix $\rho_{\hat{\Phi}} \in \mathcal{A}_R$ that satisfies the property \cite{Witten:2021unn,Chandrasekaran:2022eqq}:
\bea
\operatorname{tr}\left[\rho_{\Phi} \hat{a}\right]=e^c\langle\hat{\Phi}| \hat{a}|\hat{\Phi}\rangle\,, \quad \forall \hat{a} \in \mathcal{A}_R\,. \label{defdensitymatrixeq}
\eea
Note that this trace is defined only up to a multiplicative constant. Using this trace, we can then define a von Neumann entropy,
\bea
S_{\text{vN}}(\rho_{\hat{\Phi}}) = -\langle\hat{\Phi}| \log \rho_{\hat{\Phi}}|\hat{\Phi}\rangle+c\,. \label{entanglemententropyeq}
\eea

Factors of type I$_{d}$ and type II$_{\infty}$ have a standard form, which we will quickly review here (see \cite{Sorce:2023fdx} for more details). Let us first look at a type I$_{d}$ factor acting on a $d$-dimensional Hilbert space $\mathcal{H}$. By definition, a type~I algebra has a non-zero minimal projector $P$. We can also construct a family of pairwise orthogonal projectors $\{P_i\}$, which are equivalent to $P$, satisfying the identity $\sum_i P_i = 1$. Consider the Hilbert space $\mathcal{H}^{\prime}$ spanned by vectors $|P_i\rangle$ associated to the projectors. Then the algebra can be mapped to the tensor product $\mathcal{B}(\mathcal{H}^{\prime})\otimes 1_{P\mathcal{H}}$ by a unitary conjugation \cite{Sorce:2023fdx}. Note that the Hilbert space $\mathcal{H}^{\prime}$ has dimension $d$.
One can think of this decomposition in the following way. In a type I algebra, all minimal projectors are equivalent to each other. So the algebra splits into the product of $\mathcal{B}(\mathcal{H}^{\prime})$ and a c-number multiple of the identity operator acting in the subspace obtained by acting with \emph{some} minimal projector $1_{P\mathcal{H}}$. 

Now consider a type II$_{\infty}$ algebra $\mathcal{A}$ and let $P$ be a non-zero projector. Since the algebra is type~II, the projector $P$ will not be minimal but we can still find a set of projectors $\{P_i\}$ equivalent to $P$ that resolve the identity operator in the algebra. We identify the Hilbert space $\mathcal{H}^{\prime}$ with the span of the projectors $\{P_i\}$. The type II$_{\infty}$ algebra is then isomorphic to $\mathcal{B}(\mathcal{H}^{\prime})\otimes P\mathcal{A}P$. The difference between a type I$_d$ and a type II$_\infty$ algebra is twofold. First of all, the Hilbert space $\mathcal{H}^{\prime}$ is finite-dimensional in the case of a type I$_d$ algebra, while it is infinite-dimensional in the case of the type II algebra. Moreover, $P\mathcal{A}P$ reduces to a multiple of the identity operator in the case of a type I algebra since we have minimal projectors. 

It turns out that RMT completion provides a natural construction of a type~I$_d$ algebra starting from the type~II$_\infty$ crossed product algebra. As explained in the previous subsection, the elements of the large $N$ enlarged algebra $\mathcal{A}_R$ can be expressed as an integral over the density of states $\rho_0(E)$,
\bea
\hat{a} = \int \dd{s} \  a(s) e^{ih_Rs} = \int\dd{s} \int \dd{E} \  a(s) e^{i(E-E_0)s} P_E\,, \label{dosoperatoreq}
\eea
where $P_E$ is the projector of the right boundary Hamiltonian onto a sector of energy $E$,
\bea
P_E = \rho_0(E) |E\rangle \langle E|\,.
\eea
We have dropped the subscript $R$ on the bras and kets for brevity. Note that $P_E$ is a projector onto a single energy eigenvalue and can be formally rewritten as $\delta(H_R-E)$. Since this is an unbounded operator, it is not an element of $\mathcal{A}_R$. We can, however, construct a projector that is an element of $\mathcal{A}_{R}$ as follows:
\bea
P_{E_{i},\Delta E} = \int \dd{E} \rho_0(E) \chi_{E_{i},\Delta E}(E) \ |E\rangle \langle E|\,,
\eea
where $\chi$ is the characteristic function,
\bea
\chi_{E_{i},\Delta E}(E) = \begin{cases} 
1\,, & E \in [E_i-\Delta E,E_i+\Delta E] \,,\\
0\,, & E \notin [E_i-\Delta E,E_i+\Delta E] \,.
\end{cases}
\eea
The operator $P_{E_{i},\Delta E}$ is a spectral projection of $H_R$ onto a non-zero measure subset of the spectrum, namely the interval $[E_i - \Delta E, E_i + \Delta E]$. By the spectral theorem, all such projections are bounded operators. Since $\mathcal{A}_R$ is a von Neumann algebra, it is closed in the weak operator topology, and therefore contains all such bounded spectral projections of its operators. It follows that $P_{E_i, \Delta E} \in \mathcal{A}_R$. We can see that the projector $P_{E_{i},\Delta E}$ is not a minimal projector as we can always define another projector $P_{E_{i},\Delta E^{\prime}}$ with a spread in the energy $\Delta E^{\prime} <\Delta E$. This absence of a minimal projector is what we expect from a type II$_\infty$ algebra. 

Now build a family of projectors $\{P_{E_{i},\Delta E}\}$ by fixing 
\bea
\Delta E = \frac{1}{2\rho_0(E_0)}\,,
\eea
and choosing a sequence of evenly spaced $E_i$ (with $E=E_0$ included in the sequence) separated by $2\Delta E$, so that $\sum_i P_{E_{i},\Delta E}=1$. The resulting projectors are pairwise orthogonal. The Hilbert space $\mathcal{H}^{\prime}$ is then constructed by acting with these projectors on the TFD state,
\bea
\mathcal{H}^{\prime} = \left\{ \sum_{j \in \mathbb{Z}} \alpha_j P_{E_{j},\Delta E}|\hat{\Psi}\rangle \>\big\vert \ \forall \alpha_j \in \mathbb{C} \ \text{and} \ E_j =E_0+ 2j \Delta E \ \text{satisfying} \  E_j \geq 0\right\}\,.\label{hprimehilbertspacedef}
\eea
Here $|\hat{\Psi}\rangle$ is the state corresponding to the microcanonical TFD state in the crossed product algebra \eqref{tfdcrossedproducteq}.

The construction of such a family of non-zero projectors allows us to rewrite the enlarged algebra $\mathcal{A_R}$ in its standard form:
\bea
\mathcal{A}_R \cong \mathcal{B}(\mathcal{H}^{\prime}) \otimes P\mathcal{A}_{R}P\,,\label{standardformtypeII}
\eea
where we have chosen $P =P_{E_0,\Delta E}$. Now, we can RMT complete this description by replacing $\rho_0(E)$ with the density of states $\rho(E)$ of a random matrix theory with $\overline{\rho(E)}=\rho_0(E).$\footnote{See also \cite{Chen:2025uzn}, where the inclusion of a Poisson distribution in the context of operator algebras and third quantization was discussed.} Below, we will argue that
\bea
\mathcal{B}(\overline{\mathcal{H}^{\prime}}) \otimes \overline{P\mathcal{A}_{R}P} = \mathcal{B}(\mathcal{H}_{\text{BH}}) \otimes 1_{\overline{P\mathcal{H}}}\,, \label{typeIeq}
\eea
up to small corrections suppressed by powers of $\exp{(-S_{\text{BH}})}$. Here $\mathcal{H}_{\text{BH}}$ is the RMT completed finite-dimensional Hilbert space defined in \eqref{blackholehilbertspace}. Therefore, the RMT completed cross product algebra, after an ensemble averaging, reduces to a type I$_d$ algebra!

The result in \eqref{typeIeq} is obtained as follows: First, we have to show that the second tensor factors agree on both sides of the equation, that is, $\overline{P\mathcal{A}_{R}P} = \mathbb{C} \overline{P}$.  Before RMT completion we have
\bea
P \hat{a} P = \left(\int \dd{E} \rho_0(E) \chi_{E_0,\Delta E}(E) \ |E\rangle \langle E|\right) \hat{a} \left(\int \dd{E^{\prime}} \rho_0(E^{\prime}) \chi_{E_0,\Delta E}(E^{\prime}) \ |E^{\prime}\rangle \langle E^{\prime}|\right)\,,
\eea
where $ \hat{a} \in \mathcal{A}_{R}$. By using the general form of $\hat{a}$ \eqref{dosoperatoreq}, we find that 
\bea
P \hat{a} P = \int \dd{s} \int \dd{E} \dd{E^{\prime}} \rho_0(E)\rho_0(E^{\prime}) \chi_{E_0,\Delta E}(E) \chi_{E_0,\Delta E}(E^{\prime}) e^{i(E^{\prime}-E_0)s} \left\langle E|a(s)|E^{\prime}\right\rangle |E\rangle \langle E^{\prime}|.\ \
\eea
This expression can then be RMT completed by replacing the density of states. Taking an ensemble average of the resulting expression, we get
\bea
\overline{P \hat{a} P} = \int \dd{s}\int \dd{E} \dd{E^{\prime}} \ \overline{\rho(E)\rho(E^{\prime})} \ \chi_{E_0,\Delta E}(E) \chi_{E_0,\Delta E}(E^{\prime}) e^{i(E^{\prime}-E_0)s} \left\langle E|a(s)|E^{\prime}\right\rangle |E\rangle \langle E^{\prime}|.\ \  \label{minimalprojectoractioneq}
\eea
The presence of the characteristic functions restricts the energy values that contribute to the integral to a narrow range centered on $E_0$. This means we have to go beyond the Poisson approximation in \eqref{2ptdensityofstateseq} to evaluate the integral and instead we work with a Gaussian Unitary Ensemble (GUE),
\bea
\overline{\rho(E) \rho\left(E^{\prime}\right)}=\rho_0(E) \rho_0\left(E^{\prime}\right)+\rho_0(E) \delta\left(E-E^{\prime}\right) - \frac{\sin^2\left(\pi \rho_0(E_{\text{avg}})\left(E-E^{\prime}\right)\right)}{\pi^2 \left(E-E^{\prime}\right)^2}\,,\label{2ptdensityofstatessineeq}
\eea
where $E_{\text{avg}} =(E+E^{\prime})/2$ is the average energy. The third term on the right hand side is referred to as the \emph{sine-kernel} and encodes the level repulsion of the RMT. This term is present in all Gaussian ensembles in the $E\to E^{\prime}$ limit \cite{Mehta2004}, so the choice of a Gaussian unitary ensemble is not restrictive.

Before we continue with the computation of \eqref{minimalprojectoractioneq}, let us examine the sine kernel contribution to the averaged two-point function \eqref{2ptdensityofstatessineeq}. In particular, let us look at the case where $E_{\text{avg}} \gg |E-E^{\prime}|$. This limit corresponds to $E_0 \gg \Delta E$, so it is precisely the range of energies that contribute to the integral \eqref{minimalprojectoractioneq}. In this case, we can approximate the sine-kernel with a Gaussian function as follows:
\bea
\overline{\rho(E) \rho\left(E^{\prime}\right)}= \rho_0(E_{\text{avg}})^2\left[1-\exp\left(-\frac{\pi^2 \rho_0(E_{\text{avg}})^2\left(E-E^{\prime}\right)^2}{3}\right)\right] +\rho_0(E) \delta\left(E-E^{\prime}\right). \label{Gaussianapproximatin}
\eea
The Gaussian approximation to the sine kernel is obtained by matching \eqref{2ptdensityofstatessineeq} and \eqref{Gaussianapproximatin} up to second order in a Taylor expansion in the parameter $\rho_0(E_{\text{avg}})\left(E-E^{\prime}\right)$. Gaussian approximations to sinc-type functions are widely used in optics and signal processing to enable analytic treatment of diffraction and phase-matching problems \cite{Baghdasaryan:2022vig}. In our case, this approximation allows us to obtain closed-form expressions for integrals.

It is now easy to see how RMT corrections introduce level repulsion in the large $N$ theory. Let us ignore the delta function term in \eqref{Gaussianapproximatin} for the moment. When the energy difference is larger than $\rho_0(E_{\text{avg}})^{-1}$, the Gaussian term can be dropped, and the averaged two-point function of the density of states is equal to the product of the density of states. Thus, for large energy separations, the spectrum behaves as if it were uncorrelated. In contrast, when the energy difference drops below $\rho_0(E_{\text{avg}})^{-1}$, the averaged two-point function becomes parametrically suppressed.

Now let us return to the computation of the integral in equation \eqref{minimalprojectoractioneq}. Plugging in the averaged two-point function, we get
\bea 
\begin{aligned} 
\overline{P \hat{a} P} = \int \dd{s} \int & \dd{E} \dd{E^{\prime}}  \ \rho_0(E_{\text{avg}})^2\left[1-\exp\left(-\frac{\pi^2 \rho_0(E_{\text{avg}})^2\left(E-E^{\prime}\right)^2}{3}\right)\right]\\
&\ \chi_{E_0,\Delta E}(E) \chi_{E_0,\Delta E}(E^{\prime}) e^{i(E^{\prime}-E_0)s} \left\langle E|a(s)|E^{\prime}\right\rangle \ |E\rangle \langle E^{\prime}|\\
&\hspace{6cm}+ O(\rho_0(E_{0})^{-1})\,.
\end{aligned} 
\eea
The $O(\rho_0(E_{0})^{-1})$ correction term in the above expression comes from the delta function term in \eqref{Gaussianapproximatin} and is exponentially suppressed in the Bekenstein-Hawking entropy of the black hole, as will be clear from equation \eqref{dosbheq} below. This `contact' term in the density of states is also responsible for the exponentially suppressed plateau term in the spectral form factor.

It is convenient to change the integration variables from $(E, E^{\prime})$ to $(E_{\text{avg}}, \omega)$, where we define $\omega = E - E^{\prime}$. We now adopt a standard assumption from the RMT literature. Since $E_0 \gg \Delta E$, we assume that the matrix elements and the states depend only on the average energy $E_{\text{avg}}$. This diagonal approximation gives us
\bea 
\begin{aligned} 
\overline{P \hat{a} P} \simeq \int \dd{s} \int_{E_0-\Delta E}^{E_0+\Delta E} \dd{E_{\text{avg}}} &\ \rho_0(E_{\text{avg}})^2 e^{i(E_{\text{avg}}-E_0 )s} \left\langle E_{\text{avg}}|a(s)|E_{\text{avg}}\right\rangle  \\
&\int_{-2\widetilde{\Delta E}}^{2\widetilde{\Delta E}}  \dd{\omega} \left[1-\exp\left(-\frac{\pi^2 \rho_0(E_{\text{avg}})^2\omega^2}{3}\right)\right] |E_{\text{avg}}\rangle \langle E_{\text{avg}}| \,,
\label{labelledeq}
\end{aligned} 
\eea
where we have introduced a new quantity $\widetilde{\Delta E} =\Delta E-\left|E_{\text{avg}}-E_0\right|$ to express the above equation compactly. Evaluating the $\omega$ integral in closed form yields 
\bea 
\begin{aligned} 
\overline{P \hat{a} P} = \int \dd{s} \int_{E_0-\Delta E}^{E_0+\Delta E} \dd{E_{\text{avg}}} &\ \rho_0(E_{\text{avg}})^2 e^{i(E_{\text{avg}} -E_0)s} \left\langle E_{\text{avg}}|a(s)|E_{\text{avg}}\right\rangle  \\
&\left(4 \widetilde{\Delta E}-\frac{\sqrt{\frac{3}{\pi }} \text{erf}\left(\dfrac{2\pi  \widetilde{\Delta E} \rho_0(E_{\text{avg}}) }{\sqrt{3}}\right)}{\rho_0(E_{\text{avg}})}\right) |E_{\text{avg}}\rangle \langle E_{\text{avg}}| \,.
\end{aligned} 
\eea
The functions $\rho_0(E_{\text{avg}})$ and $\langle E_{\text{avg}}|a(s)|E_{\text{avg}}\rangle$ vary slowly over the narrow integration range, so we approximate them by their values at $E_0$ and pull them out of the $E_{\text{avg}}$ integral to obtain
\bea  
\overline{P \hat{a} P} \approx \left(\Delta E^2 \int \dd{s}  \ \rho_0(E_{0}) \left\langle E_{0}|a(s)|E_{0}\right\rangle  \right) \rho_0(E_{0}) |E_{0}\rangle \langle E_{0}| \,.
\eea
The same steps lead to the expression $\overline{P} \approx \rho_0(E_0) |E_0\rangle \langle E_0|$ for the avaraged
projector and therefore, we have $\overline{P\mathcal{A}_{R}P} \simeq \mathbb{C} \overline{P}$, as advertised.

We have shown that $P$ acts as a minimal projector after RMT completion, but our final result followed from approximating an integral by its value at a point in its domain of integration. This suggests that another projector onto an even narrower energy window,
\bea
Q = \int \dd{E} \rho(E) \chi_{E_0,\Delta E^\prime}(E) \ |E\rangle \langle E|\,, \quad \text{with} \ \Delta E^{\prime}< \Delta E \label{secondprojectoreq}
\eea
will also give us $\overline{Q\mathcal{A}_{R}Q} \approx \mathbb{C} \overline{Q}$. Clearly $Q <P$, making $P$ a non-minimal projector, but when we RMT complete and average over the projectors, the situation is different.
Let us first assume that $\Delta E^{\prime} \lesssim \Delta E$, i.e. the new energy range is only a little bit narrower than before. In this case, we can go through the same computations to find that 
\bea
\overline{Q\hat{a}Q} \approx \overline{P\hat{a}P} \implies \overline{Q} \approx \overline{P} \,.
\eea
When $\Delta E^{\prime} \ll \Delta E$, however, the behaviour changes in an interesting way. 
In this case, contributions from the two terms inside the square bracket in \eqref{labelledeq} cancel against each other, leading to a parametric suppression in the final result,
\bea
\overline{Q\hat{a}Q} \sim \left(\frac{\Delta E^{\prime}}{\Delta E}\right)^2 \overline{P \hat{a} P}\,.
\eea
Therefore, we arrive at the statement
\bea
\forall Q<P\,, \qquad \text{either} \quad \overline{Q} \approx \overline{P} \quad \text{or} \quad \overline{Q} \approx 0\,.
\eea
In this sense, $\overline{P}$ behaves as a minimal projector, as any subprojection $\overline{Q}$ gives negligible matrix elements or is approximately equal to $\overline{P}$. This is a direct consequence of the level repulsion in RMT. When the energy separation is below the characteristic scale of level repulsion in the theory, the energy levels are effectively discrete.

Now we return to \eqref{typeIeq} and show that first tensor factors also match, i.e. $\mathcal{B}(\overline{\mathcal{H}^{\prime}}) =\mathcal{B}(\mathcal{H}_{\text{BH}})$. We can construct a projector from the discrete Fourier transform of  elements of $\mathcal{B}(\mathcal{H}_{\text{BH}})$ as follows:
\bea
    \widetilde{P}_{E_i} = 2\pi\sum_{j=1}^{k} e^{-iE_i t_j}\,e^{iH_R t_j}\,, \qquad t_j= j t_1\,.
\eea
The energy resolution of $\widetilde{P}_{E_i}$ is of order $1/(k t_1)$. Since $\mathcal{H}_{\text{BH}}$ is finite dimensional, with dimension $e^{S_{\text{BH}}}$, and $t_1=O(\beta)$, this resolution matches that of the minimal projectors $P_{E_i,\Delta E}$ (we will show this in \eqref{dosbheq}). Therefore, the projectors $P_{E_i,\Delta E}$ can be reconstructed from operators in $\mathcal{B}(\mathcal{H}_{\text{BH}})$. This establishes that $\mathcal{B}(\overline{\mathcal{H}^{\prime}}) \subset \mathcal{B}(\mathcal{H}_{\text{BH}})$. Now let us look at the reverse inclusion. The right Hamiltonian can be approximated as 
\bea
h_R \approx \sum_i (E_i-E_0) \ P_{E_{i},\Delta E}\,. \label{Hamiltonianapproxiamtionprojector}
\eea
When we RMT complete the Hilbert spaces, we introduce an effective level spacing of $\Delta E$ through level repulsion, and as a result, \eqref{Hamiltonianapproxiamtionprojector} becomes an excellent approximation. This gives us $\mathcal{B}(\overline{\mathcal{H}^{\prime}}) \subset  \mathcal{B}(\mathcal{H}_{\text{BH}})$. Therefore, we have $\mathcal{B}(\overline{\mathcal{H}^{\prime}}) = \mathcal{B}(\mathcal{H}_{\text{BH}})$, concluding the proof of \eqref{typeIeq}.

Now let us revisit the computation of the dimensionality of the RMT-completed Hilbert space $\mathcal{H}_{\text{BH}}$ outlined in Section \ref{boundarymicrostatesec}. Our goal is to reformulate this calculation in the language of operator algebras. We start with the microcanonical TFD state $|\hat{\Psi}\rangle$ defined in equation \eqref{tfdcrossedproducteq}. We denote the density matrix associated to this state by $\rho_{\hat{\Psi}}$. Then for every operator $\hat{a}$ in the crossed product algebra, the trace of the operator in the state $|\hat{\Psi}\rangle$ is given by \eqref{defdensitymatrixeq}. Since we are dealing with a type II$_\infty$ algebra, the trace is non-unique with the ambiguity encoded in the constant $c$ in \eqref{defdensitymatrixeq}.
In our case, however, there is a natural way to fix this constant. The key point is that the trace on the crossed product algebra induces a trace on the factor $\mathcal{B}(\mathcal{H}^\prime)$, which in turn defines an inner product on $\mathcal{H}^\prime$. We can therefore fix the normalization by imposing the condition that the trace of the ``minimal'' projector $P$ in the microcanonical TFD state is unity,\footnote{The ambiguity in the definition of the trace arises because the decomposition of a type II$_\infty$ algebra into the standard form \eqref{standardformtypeII} is not unique (see Section 3.6 of \cite{Witten:2021jzq}). In general, one may choose any non-minimal projector $P$ to obtain such a decomposition and then impose the condition \eqref{traceconditioneq} to fix the trace, leading to multiple possible choices for the constant $c$. However, the RMT completion produces a minimal projector, which in turn provides a canonical choice for $c$.}
\bea
\mathrm{tr}[\rho_{\hat{\Psi}}P] = 1\,. \label{traceconditioneq}
\eea
Using the definition \eqref{tfdcrossedproducteq} of the state $|\hat{\Psi}\rangle$, we have
\bea
\mathrm{tr}[\rho_{\hat{\Psi}} P] = e^c \int_{-\infty}^{\infty} \dd{x}\ (f(x))^2\langle x| \langle\Psi| P|\Psi\rangle |x\rangle\,.
\eea
We now express the projector $P$ in terms of its Fourier representation,
\bea
P = \int \dd{s}\ b(s) e^{is(X+\hat{h}+E_0)}\,,
\eea
where $b(s)$ are the Fourier coefficients that implement the finite energy range of the characteristic function $\chi_{E_0,\Delta E}$. Substituting this into the trace gives
\bea
\mathrm{tr}[\rho_{\hat{\Psi}} P] = e^c \int_{-\infty}^{\infty} \dd{x} \int \dd{s}\  b(s) (f(x))^2 \ \langle\Psi| e^{is(x+\hat{h}+E_0)}|\Psi\rangle\,.
\eea
Since $\hat{h}|\Psi\rangle = 0$, the matrix element simplifies, and we obtain
\bea
\mathrm{tr}[\rho_{\hat{\Psi}} P] =   e^c \int_{-\infty}^{\infty} \dd{x}\ (f(x))^2 \int \dd{s}\ b(s) e^{is(x+E_0)}=   e^c \int_{-\infty}^{\infty} \dd{x}\ (f(x))^2\chi_{E_0,\Delta E}(x+E_0)\,.
\eea
Here the integral over $s$ is the Fourier transform of $b(s)$, which reduces to the characteristic function $\chi$. Since $f(x)$ is slowly varying on the scale of $\Delta E$, we can pull it out of the integral and evaluate it at the center. We then obtain the normalization condition
\bea
\rho_0(E_0) \simeq e^c (f(0))^2 \,.
\eea
The normalization of the density of states can also be read off from \eqref{microtfdeq}, giving 
\bea
\rho_0(E_0) \simeq e^{S_{\text{BH}}}(f(0))^2\,. \label{dosbheq}
\eea
Comparing these expressions, we find
\bea
c \simeq S_{\text{BH}}\,.
\label{clabel}
\eea

As in the previous subsection, we define time-shifted thermofield states by evolving with the right Hamiltonian,
\bea
|\hat{\Psi}(t)\rangle \equiv e^{i\hat{h}t} e^{iXt}|\hat{\Psi}\rangle\,.
\eea
We can associate density matrices $\rho_{\hat{\Psi}(t)}$ to these states. Using the definition \eqref{defdensitymatrixeq} together with the cyclicity of the trace, we find that 
\bea
\rho_{\hat{\Psi}(t)} = e^{i\hat{h}t} e^{iXt} \rho_{\hat{\Psi}}e^{-i\hat{h}t} e^{-iXt}\,.
\eea
Now consider the quantity
\bea
\mathrm{tr}\left[\rho_{\hat{\Psi}(t_1)}\rho_{\hat{\Psi}(t_2)}\right]\,,
\eea
which equals the square of the overlap in equation \eqref{overlapsquaredeq}. From the definition of the trace, we find that
\bea
\mathrm{tr}\left[\rho_{\hat{\Psi}(t_1)}\rho_{\hat{\Psi}(t_2)}\right]= e^c\langle\hat{\Psi}| e^{-i h_R t_1} \rho_{\hat{\Psi}} e^{i h_R t_1} |\hat{\Psi}\rangle\,.
\eea
Expanding the right Hamiltonian in terms of the density of states, we get
\bea
e^{-i h_R t_1} \rho_{\hat{\Psi}} e^{i h_R t_1} = \int \dd{E}\dd{E^{\prime}} \ \rho_0(E) \rho_0(E^{\prime}) \  e^{i (t_2-t_1)(E-E^{\prime})} \langle E| \rho_{\hat{\Psi}} |E^{\prime} \rangle \ | E\rangle  \langle E^{\prime} |\,.
\eea
Now, let us take an average of this quantity in the RMT completion. We can use the Poisson distribution \eqref{2ptdensityofstateseq} to get the connected component
\bea
\left.\overline{\mathrm{tr}\left[\rho_{\hat{\Psi}(t_1)}\rho_{\hat{\Psi}(t_2)}\right]}\right|_c = e^c\langle\hat{\Psi}| \rho_{\hat{\Psi}} |\hat{\Psi}\rangle\,.
\eea
Then by the definition of the von Neumann entropy \eqref{entanglemententropyeq}, we have
\bea
\left.\overline{\mathrm{tr}\left[\rho_{\hat{\Psi}(t_1)}\rho_{\hat{\Psi}(t_2)}\right]}\right|_c = e^{-S_{\text{vN}}(\rho_{\hat{\Psi}})}\,.
\eea
The von Neumann entropy can be shown to be equal to \cite{Witten:2021unn,Chandrasekaran:2022eqq}
\bea
S_{\text{vN}}(\rho_{\hat{\Psi}}) = c-\langle\widehat{\Psi}| \log\left|f\left(h_R\right)\right|^2|\widehat{\Psi}\rangle = S_{\text{BH}}+S_\text{log}\,,\label{bhentropy}
\eea
where we have used \eqref{clabel} to fix the value of $c$ in the final step.
The second term $S_\text{log}$ is the ``fluctuation entropy,'' which captures the universal logarithmic corrections to the black hole entropy \cite{Das:2001ic}. The dimensionality of the Hilbert space $\overline{H_{\text{BH}}}$ is given by $d=e^{S_{\text{BH}}+S_\text{log}}$. 

Holographic duality allows us to carry out a similar construction in the bulk. In the case of an eternal black hole, the right algebra corresponds to the algebra of a massive scalar field on the right wedge of the spacetime. Instead of the thermofield state, we have the Hartle-Hawking state. The modular Hamiltonian generates boosts inside the wedge. The analogue of the right CFT Hamiltonian will be played by the right ADM Hamiltonian. Going through all the steps above, we would arrive at a type I$_d$ algebra in the bulk.

\subsection{Complexity as a Probe of the Emergent Geometry}
Once the microstates of the theory have been constructed, it is natural to ask when the low-energy semiclassical description breaks down. In the language of time-shifted TFD states, this corresponds to where the non-perturbative RMT corrections to the density of states dominate. We note from \eqref{secondoverlapeq} that the overlap of the time-shifted TFD states is the spectral form factor,
\bea
\overline{\left|\left\langle \text{TFD}(t_1)|\text{TFD}(t_2)\right\rangle \right|^2} = \text{SFF}(|t_1-t_2|,\beta)\,.
\eea
The spectral form factor of RMT has a distinct behaviour \cite{Cotler:2016fpe}. Plotted as a function of time, there is an initial dip, followed by a ramp and plateau. During the dip, the SFF decays for a long time, and only when $t\sim e^{O(S_{\text{BH}})}$ do the ramp and plateau become visible. The late-time ramp and the plateau are a result of the sine-kernel and the Dirac delta terms dominating the averaged two-point function of the density of states \eqref{2ptdensityofstatessineeq}. At this exponential timescale the semiclassical description breaks down.\footnote{As emphasized in the Introduction, the semiclassical description is here taken to include perturbative quantum gravity corrections but non-perturbative corrections are neglected. It is important to distinguish this from the small but non-zero $G_N$ limit of the full theory, where non-perturbative corrections are exponentially suppressed but nevertheless always present.}

A natural way to reframe these observations is through the notion of complexity. In particular, consider Krylov complexity \cite{Parker:2018yvk}, which assigns a ``size'' to an operator under Hamiltonian evolution. In chaotic systems, operator complexity exhibits a generic pattern, growing linearly with time for a long period, and saturating at times of exponential order in the system's entropy \cite{Barbon:2019wsy}. The ramp and plateau in the SFF are responsible for this saturation of complexity (see \cite{Iliesiu:2021ari} for details). From this perspective, the semiclassical description breaks down if we probe the black hole spacetime with an operator whose complexity is of exponential order in the Bekenstein-Hawking entropy. See \cite{Akers:2022qdl,Krishnan:2023fnt,Balasubramanian:2026azk} for related discussions.
This idea can be made very explicit by using the Complexity=Volume (CV) prescription \cite{Susskind:2014rva,Stanford:2014jda}, where bulk complexity is defined as the volume of a spacelike extremal codimension-one surface passing through the interior of the black hole. In two-dimensional spacetimes, there are bulk wormhole corrections to the semiclassical volume computations when the boundary anchoring time of the surface reaches $e^{O(S_{\text{BH}})}$ \cite{Iliesiu:2021ari}, and this leads to the saturation of complexity. Similar behaviour is found in higher dimensions \cite{Balasubramanian:2022gmo,Gautason:2025ryg}. This breakdown of the semiclassical picture has important consequences. The region where the semiclassical volume calculation fails is precisely the region where certain \emph{trapped extremal surfaces} appear \cite{Mohan:2025acj}. These surfaces lead to null geodesic incompleteness in the interior and signal the presence of the black hole singularity. The corrections responsible for the saturation of holographic complexity must therefore also modify the volume computation that gives rise to these pathological surfaces, hinting at a possible resolution of black hole singularities.

A puzzling feature of the geometry probed by an operator is that it appears to be highly \emph{observable dependent}. At first glance, this may seem troubling but it lies at the heart of the principle of black hole complementarity \cite{Susskind:1993if}. When a quantum black hole is probed with a “simple operator”, the resulting spacetime looks like the classical geometry. There is an event horizon, and an infalling observer experiences a smooth infall. However, when the black hole is probed with a “complex operator”, the semiclassical limit breaks down. One may encounter a wormhole or a baby universe emission. The crucial point is that these different observable-dependent descriptions are complementary — each measurement probes a different operational regime.

\section{Discussion}
\label{discussionsec}
Operator algebras are useful for the study of quantum gravity and the emergence of spacetime as these techniques eliminate the need for (Euclidean) path integrals and their saddle-point approximations, and do not rely on formulating the theory in terms of a Lagrangian. In this framework, Hilbert spaces emerge as representations of operator algebras. This perspective is particularly useful, since treating Hilbert spaces as fundamental objects leads to conceptual puzzles such as the factorisation problem (see, for example, \cite{Harlow:2018tqv,Boruch:2024kvv}).

A remarkable feature of general relativity is that gravity is encoded in the curvature of spacetime through the equivalence principle and Einstein's field equations. Our analysis in Section \ref{geometrysection} shows that both the equivalence principle and the field equations follow from the existence of multiple local Rindler algebras and locally stationary states, along with a well-defined large-mass limit for the matter field correlation functions. This provides a characterization of gravity purely in terms of operator algebras, which are the natural building blocks of quantum mechanics. The reformulation of geometry carries philosophical implications, as gravity is no longer viewed as fundamental but instead emerges from the operator algebras of \emph{matter fields}. 
An important further step is to extend our analysis to more realistic matter fields of the kind present in our universe and to understand how the constructions in this paper generalize to that setting. 

We would like to emphasize that, in Section \ref{geometrysection}, we have not assumed the existence of a boundary dual. In particular, the existence of a smooth $d$-dimensional spacetime in the boundary theory would significantly simplify the reconstruction of the bulk metric and curvature tensors, since boundary points could then be used to localize bulk points as in, for example, \cite{Engelhardt:2016wgb}. One could then directly use the smooth differentiable structure of the boundary to extract the metric and curvature tensors \cite{Jiang:2024hjz}. Our approach deliberately avoids making this assumption.

In Section \ref{typeIsection}, we showed how a type I algebra can be obtained from a type III algebra by a random matrix theory completion of the crossed product algebra. There is mounting evidence for randomness and the necessity of averaging in operator-algebraic descriptions of quantum gravity, see e.g. \cite{Liu:2025cml,Liu:2025ikq,Kudler-Flam:2025cki}. The RMT construction restores the discreteness of the spectrum that is lost in the large $N$ limit and naturally leads to chaos and thermalization. Once these effects are incorporated, the completed algebra also captures non-perturbative contributions from bulk wormholes. An important aspect of our RMT implementation is that it does not call for a modification of the standard AdS/CFT correspondence, which has passed numerous precision tests. We use the effective RMT description to capture the large $N$ semiclassical limit of the theory, which is a standard strategy for understanding generic chaotic systems and is not restricted to gravity.

\acknowledgments
We would like to thank Rahul Poddar for useful discussions. VM is supported by a doctoral grant from the University of Iceland Science Park.

\appendix

\section{Basic Definitions and von Neumann Algebras}
\label{reviewapp}
In this appendix, we provide a brief review of some key definitions and properties of von Neumann algebras. We recommend \cite{Sorce:2023fdx,Haag1992} for a more comprehensive treatment.

Consider a Hilbert space $\mathcal{H}$. We will denote the algebra of all bounded local operators acting on this Hilbert space by $\mathcal{B}(\mathcal{H})$. A \emph{$*$-subalgebra} is a subalgebra of $\mathcal{B}(\mathcal{H})$ that is closed under scalar multiplication, operator multiplication, operator addition, and adjoints. It also contains the identity element. The commutant algebra $\mathcal{M}^{\prime}$ is the set of all operators in $\mathcal{B}(\mathcal{H})$ that commute with every element of a subalgebra $\mathcal{M}$. A \emph{von Neumann algebra} $\mathcal{M}$ is a subalgebra of $\mathcal{B}(\mathcal{H})$ that is equal to its own double commutant, that is, $\mathcal{M}=\mathcal{M}^{\prime \prime}$. A von Neumann algebra $\mathcal{M}$ is a \emph{factor} if the center $\mathcal{Z} \equiv \mathcal{M} \cap \mathcal{M}^{\prime}$ is composed of scalar multiples of the identity operator.

A function $\omega$ from an algebra $\mathcal{M}$ to the complex numbers is called a \emph{state} if it is 
\begin{enumerate}
  \item linear : $\omega(\alpha a + \beta b) =\alpha  \omega( a)+\beta\omega(b)$, for all $a,b \in \mathcal{M}$.
  \item positive : $\omega(a^{*}a)\geq 0,  \ \forall a \in \mathcal{M} $.
  \item normalized : $\omega(1) =1$.
\end{enumerate}

An orthogonal projector $P$, or simply a \emph{projector}, is an element of a von Neumann algebra $\mathcal{M}$ with the properties $P=P^2=P^{*}$, where $P^{*}$ is the adjoint operator. A pairwise orthogonal set of projectors $\{P_i\}$ satisfies $P_iP_j = \delta_{ij} P_j$ for all $i,j$.

Two projectors $P$ and $Q$ are equivalent if and only if there exists an operator $V\in \mathcal{M}$ with $VV^*=P$ and $V^*V=Q$. We denote this relation by $P\sim Q$. We write $P>Q$ if the projectors $P$ and $Q$ are not equivalent and $P\mathcal{H} \supset Q\mathcal{H}$. We then refer to $Q$ as a proper subprojector of $P$.

A projector $P$ is finite if every non-zero proper subprojector $Q< P$ is inequivalent to $P$. A non-zero projector is \emph{minimal} if $P\mathcal{M}P = \mathbb{C}P$. Using this definition, we can classify von Neumann algebras into the following categories
\begin{enumerate}
 \item \textbf{Type I} : contains a non-zero minimal projector $P$.
 \item \textbf{Type II} : contains non-zero finite projectors, but no non-zero minimal projectors.
 \item \textbf{Type III} : contains no non-zero finite projectors.
\end{enumerate}

We call a vector $|\omega\rangle \in \mathcal{H}$ \emph{cyclic} if the set of $a|\omega\rangle$, for all $a \in \mathcal{M}$, is dense in $\mathcal{H}$. The vector is \emph{separating} if $a|\omega\rangle =0$ iff $a=0$. If $|\omega\rangle$ is cyclic and separating, then Tomita-Takesaki theory guarantees the existence of two operators $\Delta$ and $J$ \cite{Takesaki:1970aki}. The first operator $\Delta$ is called the \emph{modular operator} while the second operator is called \emph{modular conjugation}.

The modular operator generates modular flows in the von Neumann algebra:
\bea
\Delta^{it} \mathcal{M} \Delta^{-it} = \mathcal{M}\,, \qquad \forall t \in \mathbb{R}\,. 
\eea
The modular Hamiltonian $\hat{h}$ is related to the modular operator through the relation
\bea
\Delta = e^{-\hat{h}}\,.
\eea
The modular conjugation takes elements of the algebra to its commutant:
\bea
J\mathcal{M}J = \mathcal{M}^{\prime}\,.
\eea
\begin{figure}
\centering
\includegraphics[width=0.9\linewidth]{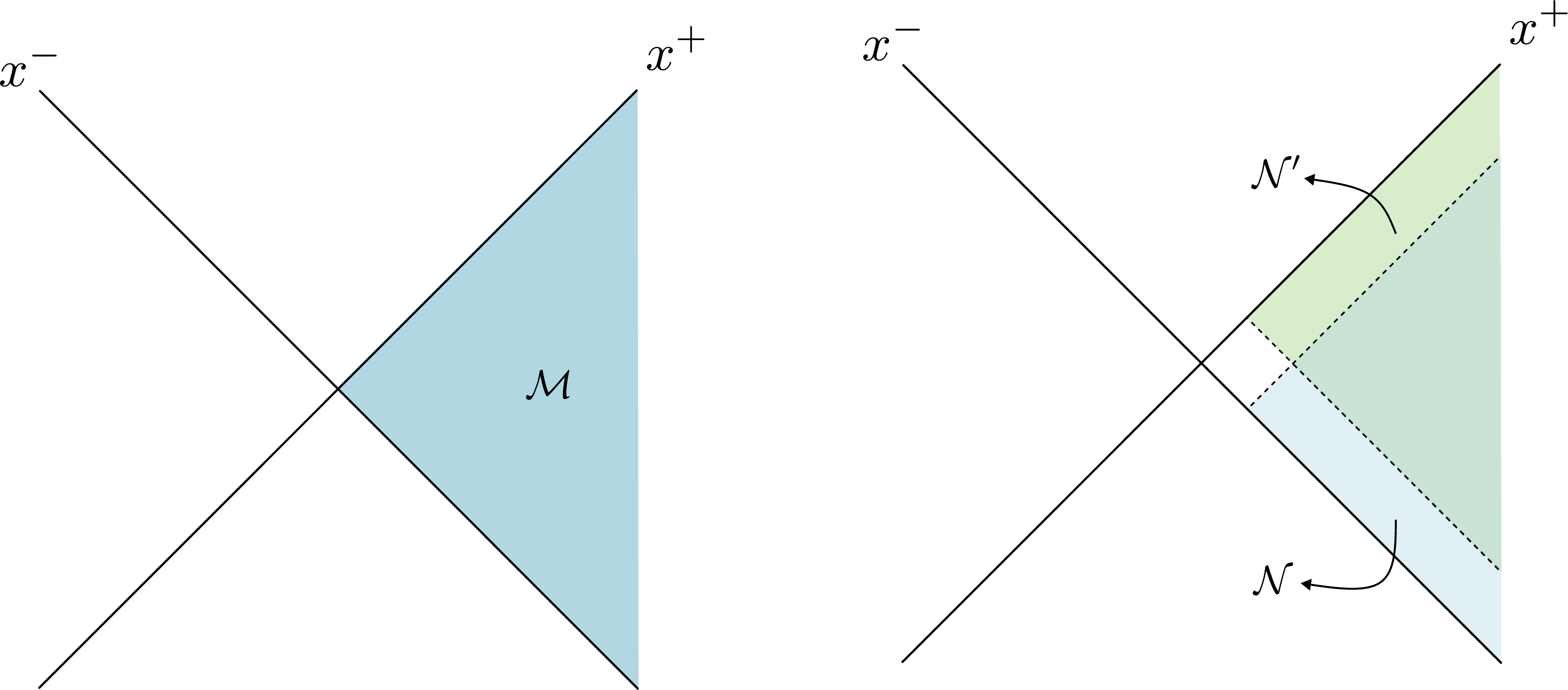}
\caption{\small The figure on the left shows the entire right wedge algebra $\mathcal{M}$. On the right are two subalgebras $\mathcal{N},\mathcal{N}^{\prime}$ obtained by shifting the right wedge along the null directions.}
\label{wedgesfig}
\end{figure}The notion of half-sided modular inclusions was introduced by Wiesbrock \cite{Wiesbrock:1992mg}. Consider a subalgebra $\mathcal{N} \subset \mathcal{M}$ for a von Neumann algebra $\mathcal{M}$ with a common cyclic separating vector $|\omega\rangle$. Let us also assume that the subalgebra is preserved under the modular flow of $\mathcal{M}$:
\bea
\Delta_{\mathcal{M}}^{i t} \mathcal{N} \Delta_{\mathcal{M}}^{-i t} \subset \mathcal{N} \quad \forall  \ t \leq 0 .
\eea
Half-sided modular inclusions are generated by a one-parameter family of operators given by
\bea
U(s) = e^{-iGs}\,,
\eea
where
\bea
2\pi G = \hat{h}_{\mathcal{M}} -\hat{h}_{\mathcal{N}}\geq 0\,.
\eea
Here $\hat{h}_{\mathcal{M},\mathcal{N}}$ are the modular Hamiltonians of the two algebras. Now we will study an example to see how the boosts, along with the half-sided modular translations $U(s)$, can be used to obtain the generators of the spacetime.

Consider a $1+1$-dimensional Minkowski spacetime written in null coordinates $x^{\pm}$. Let us take a massive scalar field whose support is restricted to the right Rindler wedge defined by $x^{+}\geq 0$ and $x^{-}\leq 0$, and denote the corresponding algebra of operators by $\mathcal{M}$. Now introduce a shifted wedge algebra $\mathcal{N}$, consisting of operators supported in the region $x^{+}\geq 0$ and $x^{-}\leq -1$ (see Figure \ref{wedgesfig}). One can then show that the modular Hamiltonian associated with $\mathcal{M}$ generates the Rindler boost $K$ in the right wedge, while the half-sided modular translations associated with the inclusion $\mathcal{N}\subset\mathcal{M}$ are given by \cite{Leutheusser:2021frk}
\bea
U_+(s)= e^{-isP^+}\,,
\eea
where $P^{+}$ is the generator of translations along $x^{+}$.

If we choose another subalgebra $\mathcal{N}^{\prime}$ composed of operators with support in the region $x^{+}\geq1$ and $x^-\leq 0$ (see Figure \ref{wedgesfig}), then we will end up with another half-sided modular translation
\bea
U_-(s)= e^{-isP^-}\,.
\eea
The generators $\{K,P^{\pm}\}$ close to form the two-dimensional Poincar\'{e} algebra. In Appendix \ref{poincareappendix}, we demonstrate that the operator algebraic construction of the Poincar\'{e} algebra generalizes to higher dimensions.

\section{Construction of the Poincar\'{e} Algebra}
\label{poincareappendix}
Let us show how the Poincar\'{e} generators of a globally flat spacetime can be constructed from operators restricted to multiple Rindler wedges. Consider a massive scalar field living on a $d+1$-dimensional spacetime. We will use the Cartesian coordinates $(t,x_1,x_2,...)$ to label the points on the spacetime.

Now let us look at the operators restricted to a Rindler wedge $\mathcal{R}_i = \{x_i>|t|\}$. As reviewed in Appendix \ref{reviewapp}, we can construct the generators of the null translations $t\pm x_i$ using the half-sided modular translations $U_{\pm}(s)$. This then gives us
\bea
e^{-i s X_0} = U_+(s/2)U_-(s/2)\,,\qquad \qquad
e^{-i s X_i} = U_+(s/2)U_-(-s/2)\,,
\eea
where $X_0$ generates time translations, while $X_i$ generates translations along the spatial direction. By repeating this construction with a different wedge algebra $\mathcal{R}_j$, we can similarly obtain a generator along another spatial direction. In this way, we build a set of translation generators, denoted by $\{X_j\}$. From this set, we can select $d$ mutually orthogonal spatial generators by imposing the condition
\bea 
\lim_{s_1,s_2\to 0}\frac{\partial^2}{\partial s_1 \partial s_2 }\sigma_{jk}(i_{s_1},i_{s_2}) =\delta_{jk}\,,\label{orthogonalityeq} 
\eea
where $\sigma_{jk}(i_{s_1},i_{s_2})$ is the Synge world function extracted from the correlator
\bea 
\bra \omega | \left(e^{is_1X_j}\mathcal{O}_ie^{-is_1 X_j}\right) \left(e^{is_2X_k}\mathcal{O}_ie^{-is_2X_k}\right) |\omega \ket\equiv G_{jk}(i_{s_1},i_{s_2})\,. 
\eea
Here $\mathcal{O}_i$ is an operator that lies in both $\mathcal{W}_j$ and $\mathcal{W}_k$. We have used \eqref{metricfromsigmaeq} to arrive at the orthogonality condition \eqref{orthogonalityeq}.

Next, let us denote by $K_j$ the boost operator associated with the wedges of each of these $d$ spatial translations. Together, the boosts $K_j$, the mutually orthogonal spatial translations $X_j$, and the time translation generator $X_0$ furnish the boosts and translations of the local Poincar\'{e} algebra. Using the commutation relations
\bea
[K_j, K_k] = -i \epsilon_{jkl} J_l,,
\eea
we can then obtain the rotation generators $J_k$, thereby recovering the full Poincar\'{e} algebra.

\bibliographystyle{JHEP}
\bibliography{refs.bib}

\end{document}